\documentclass[12pt,preprint]{aastex}
\usepackage{graphicx,amssymb,amsmath,times}
\shorttitle{Cascades from GRBs}
\shortauthors{IceCube Collab.}
\slugcomment{\today}

\begin{document}

\title{Search for neutrino-induced cascades from gamma-ray bursts with AMANDA}

\author{
IceCube Collaboration:
A.~Achterberg\altaffilmark{1},
M.~Ackermann\altaffilmark{2},
J.~Adams\altaffilmark{3},
J.~Ahrens\altaffilmark{4},
K.~Andeen\altaffilmark{5},
J.~Auffenberg\altaffilmark{6},
J.~N.~Bahcall\altaffilmark{7,a},
X.~Bai\altaffilmark{8},
B.~Baret\altaffilmark{9},
S.~W.~Barwick\altaffilmark{10}
% ...
}

% ...
\altaffiltext{1}{Dept.~of Physics and Astronomy, Utrecht University/SRON,
  NL-3584 CC Utrecht, The Netherlands}
\altaffiltext{2}{DESY, D-15735 Zeuthen, Germany}
\altaffiltext{3}{Dept.~of Physics and Astronomy, University of Canterbury, Private Bag 4800, Christchurch, New Zealand}
\altaffiltext{4}{Institute of Physics, University of Mainz, Staudinger Weg 7, D-55099 Mainz, Germany}
\altaffiltext{5}{Dept.~of Physics, University of Wisconsin, Madison, WI 53706, USA}
\altaffiltext{6}{Dept.~of Physics, University of Wuppertal, D-42119 Wuppertal, Germany}
\altaffiltext{7}{Institute for Advanced Study, Princeton, NJ 08540, USA}
\altaffiltext{8}{Bartol Research Institute and Department of Physics and Astronomy, University of Delaware, Newark, DE 19716, USA}
\altaffiltext{9}{Vrije Universiteit Brussel, Dienst ELEM, B-1050 Brussels, Belgium}
\altaffiltext{10}{Dept.~of Physics and Astronomy, University of California, Irvine, CA 92697, USA}
% ...
\altaffiltext{a}{Deceased}
% ...

\begin{abstract}
Using the neutrino telescope AMANDA-II, we have conducted two analyses
searching for neutrino-induced cascades from gamma-ray bursts.  No
evidence of astrophysical neutrinos was found, and limits are presented
for several models. We also present neutrino effective areas which allow
the calculation of limits for any neutrino production model. The first
analysis  looked for a statistical excess of events within a sliding
window of 1 or 100 seconds (for short and long burst classes,
respectively) during the years 2001-2003. The resulting upper limit on
the diffuse flux normalization times $E^2$ for the Waxman-Bahcall model at 1
PeV is 1.6$\times$10$^{-6}$~GeV~cm$^{-2}$~s$^{-1}$~sr$^{-1}$ (a factor of 120 above the theoretical prediction). For this search 
90\% of the neutrinos would fall in the energy range 50~TeV to 7~PeV.
The second analysis looked for neutrino-induced cascades in coincidence
 with 73 bursts detected by BATSE in the year 2000. The resulting upper
limit on the diffuse flux normalization times $E^2$, also at 1 PeV, is 
1.5$\times$10$^{-6}$~GeV~cm$^{-2}$~s$^{-1}$~sr$^{-1}$ (a factor of 110 above the theoretical prediction) for the same energy
range. The neutrino-induced cascade channel is complementary to the up-going
muon channel.  We comment on its advantages for searches of neutrinos from GRBs
and its future use with IceCube.

\end{abstract}

\keywords{Gamma-Ray Burst, Neutrinos, Neutrino Telescopes}

\section{Introduction}

Gamma-ray bursts (GRBs) have been proposed as one of the most plausible sources
of ultra-high energy cosmic rays \citep{wbcr,crGRB2}. In addition to being a
major advance in neutrino astronomy, detection of high energy neutrinos
from a burst would provide corroborating evidence for the acceleration of
ultra-high energy cosmic rays within GRBs.  

AMANDA-II \citep{ama:nature-pub}, the final configuration of the Antarctic
Muon And Neutrino Detector Array, is located  at the South Pole.  It was
commissioned in the year 2000 and consists of a total of 677 optical modules.
Each module contains a photomultiplier tube and supporting hardware inside a
glass pressure sphere.  These are arranged on 19 strings frozen into the ice,
with the sensors at depths ranging from 1500 m to 2000 m in a cylinder of
100~m radius. The optical modules indirectly detect neutrinos by measuring the
Cherenkov light from secondary charged particles produced in neutrino-nucleon
interactions. AMANDA is being integrated into the IceCube detector which is
currently under construction.

Searches for neutrino-induced muons in
coincidence with GRBs have been performed with the AMANDA detector for the
years 1997-2003
\citep{GRB-muon,icrc05:GRB,icrc05:GRBb,phd:hardtke,phd:bay}. Cascades, which
are electromagnetic and hadronic particle showers, provide a 
complementary channel to muon detection \citep{ama2:cascades}. This paper
presents two analyses which have searched for neutrino-induced cascade signals
from GRBs.  In the \emph{rolling} search, 3 years (2001-2003) of AMANDA-II
data were scanned for a clustering of signal events in time.  In the
\emph{triggered} search, AMANDA-II data were analyzed for a neutrino signal in
temporal coincidence with 73 bursts reported by the Burst and Transient Source
Experiment, BATSE \citep{batse}, during the year 2000.

Compared to AMANDA cascade analyses, neutrino-induced muon searches have
 higher overall event rates because the muon's long range allows detection 
 even if it is produced far outside the detector, while a cascade has 
 to happen at least partially within the detector array. Muons can also use
 directional constraints to reduce background because their linear, track-like
 shape gives them much better pointing resolution. This allows the identification of muons originating from up-going neutrinos, as these are the only known particles to propagate through the Earth. 


However, these disadvantages are balanced by several arguments in favor of
cascades. Since cascades are topologically distinct from AMANDA's primary
background of down-going atmospheric muons, it is not necessary to use the
Earth as a filter as in the case of muons. Hence, cascade analyses have full
sky  (4$\pi$~sr) coverage, as opposed to 2$\pi$~sr for muon analyses. This
doubles the number of bursts that can be studied by a single detector. For the
triggered analysis, this number is more than doubled, since bursts which do 
not have good directional localization based on satellite information can
still be used in the cascade search. Additionally, the energy resolution for
cascades is better than that for muons because of the calorimeter-like
energy deposition in the detector. For cascades produced via charged current
channels which produce only showers ($\nu_{e}$ and $\nu_{\tau}$) the energy of
the final state can be completely measured. Finally, on average the cascade
energy is more closely correlated to that of its parent neutrino than for
muons because for muons the interaction vertex is typically in an unknown
place outside of the detector. 

While neutrino-induced muon tracks are only caused by charged current
$\nu_{\mu}$ interactions, cascades can be produced by interactions of all 3
neutrino flavors.  Processes producing cascade signatures include $\nu_{x}+N$
neutral current interactions of any neutrino flavor, $\nu_{e}+N$ charged
current, $\bar{\nu}_{e}+e^-$ around 6.3 PeV (the Glashow Resonance) and
$\nu_{\tau}+N$ charged current interactions. The last case results in 
isolated cascade-like events when the  $\tau$ decays into an electron
($\sim$18\% branching ratio) or into mesons ($\sim$64\% branching ratio)
and the $\tau$ energy is below $\sim$100~TeV \citep{pdg}. The decay length of
a $\tau$ with an energy of 100 TeV is approximately 5~m, so the showers
produced by the neutrino interaction and by the $\tau$  decay cannot be
spatially resolved by AMANDA.  For neutrinos above 100 TeV, topological
searches can be used to detect $\nu_\tau$ \citep{doublebang}, but in the
analyses presented here we optimize for the search of isolated cascades and
ignore other $\nu_\tau$ event topologies.  Charged current $\nu_{\mu}$
interactions can produce cascades in addition to tracks, but this channel is
ignored in these analyses in favor of cascades which are not contaminated by
track-like signatures. 

\section{Neutrinos from Gamma-ray bursts}

It is believed that gamma rays produced by GRBs originate from electrons
accelerated in internal shock waves associated with relativistic jets
(with a bulk Lorentz boost $\Gamma$ of 100-1000) \citep{elecaccel1,elecaccel2}. These gamma rays have
energies ranging from 10 keV to 10 MeV or more. The gamma-ray spectrum can be
generically described as a broken power law, with a softer spectrum above a
break energy which is typically 30~keV-1~MeV. Gamma-ray bursts can last
anywhere from a few milliseconds to around 1000~s. The distribution
(as observed by BATSE) of durations is bimodal. For the puposes of these analyses, we define as \textit{short}
bursts those that last less than 2~s and as \textit{long} bursts those that
last longer than 2~s \citep{batse}. Other types of bursts have been proposed, but the searches presented here do not apply to these classes. Reviews of the observational
and theoretical status of gamma-ray bursts may be found in \citet{GRBreview}
and \citet{piranreview}.

If protons and/or nuclei are also accelerated in the jets, then
high energy (TeV-PeV) neutrinos are produced via the
process \citep{wb}:

\begin{equation}
\label{eqn:pgamma}
p+\gamma \rightarrow \Delta^+ \rightarrow \pi^+[+n] \rightarrow \nu_\mu +
\mu^+ \rightarrow \nu_\mu + e^+ + \bar{\nu}_{\mu} + \nu_{e}.
\end{equation}

The kinematics of this reaction are such that the average
energy of each neutrino is approximately the same, so the neutrino flavor ratio
$\nu_{e}$:$\nu_{\mu}$:$\nu_{\tau}$ is 1:2:0 at the source. Taking into account 
neutrino oscillations, the flavor ratio observed at Earth is 1:1:1
\citep{athar}. However, \citet{kashti} point out that at energies greater than
$\sim$~1~PeV, the $\mu^{+}$ in Equation~(\ref{eqn:pgamma}) loses energy
through synchrotron radiation before decaying. This effect changes the source 
neutrino flavor ratio from 1:2:0 to 0:1:0 as energy increases, leading to a
ratio at Earth of 1:1.8:1.8 at high energies for the Waxman-Bahcall neutrino
spectrum.

Even at energies where the flavor ratio is 1:1:1, the $\nu$:$\overline{\nu}$
ratio is not 1:1. This 
is because neutrinos are produced via the p$\gamma$ interaction. At the source
the neutrino flavor ratio (excluding antineutrinos) is 1:1:0 and the
antineutrino flavor ratio is 0:1:0. After taking into account preferred values
of mixing angles \citep{oscillations} for neutrino oscillations the flux
ratios at Earth are 0.8:0.6:0.6 and 0.2:0.4:0.4 for neutrinos and
antineutrinos respectively. The $\nu$:$\overline{\nu}$ flux ratio is relevant
in the calculation of the total number of expected events by the detector.

TeV-PeV neutrinos are expected to be simultaneous with prompt gamma-ray
emission. The neutrino spectrum is described by a broken power law. For both 
searches presented in this paper we will use the \citet{wb} broken power law
spectrum as a reference hypothesis and to optimize our data selection
criteria. This spectrum is:

\begin{equation}
  \label{eqn:wb}
  \frac{\mathrm{d}\Phi_\nu}{\mathrm{d}E} = A \left\{
  \begin{array}{lr}
    E^{-1}/E_{\mathrm b} & E < E_{\mathrm b} \\
    E^{-2} & E_{\mathrm b} < E < E_{\pi} \\
    E^{-4}E_\pi^2 & E > E_{\pi} 
  \end{array}
\right. ,
\end{equation}
where $A$ is the flux normalization, $E_{\mathrm{b}}$ is the break energy
corresponding to the break in the parent photon spectrum and $E_\pi$ is the
energy break due to pion energy losses. Following \citet{wb} and \citet{wb02}
we set $E_{\mathrm b}$=100 TeV, $E_{\pi}$=10 PeV and $A$=1.3$\times 
10^{-8}$~GeV~cm$^{-2}$~s$^{-1}$~sr$^{-1}$ at the Earth for all neutrino
flavors combined. In reality, each GRB is unique and the spectral shape and
normalization of individual GRBs may vary significantly from this assumed
``typical'' spectrum \citep{guetta,icrc05:GRB}. The rolling search, however,
is conducted independent of external triggers. This frees the search from
detector selection effects introduced by the gamma-ray satellites, but makes
optimizing on an averaged spectrum the only viable option. For the triggered
analysis we have chosen to optimize the selection criteria with the mean
spectrum as well. Also, selection criteria optimization is not strongly
dependent on the exact shape of the signal spectrum.

Newer models update the Waxman-Bahcall model with current
knowledge. \citet{murase} have performed a detailed simulation of neutrino
production in internal shocks in GRBs. The authors use several models
for the redshift distribution of GRBs, e.g. one assumption is that the
(long duration) GRB rate follows the star formation rate. They vary several
parameters, such as spectral hardness, to reflect current
unknowns. In this paper we present limits on the Murase-Nagataki
model. \citet{guetta} have improved on Waxman-Bahcall with a phenomenological
approach. They have used information specific to bursts reported by the BATSE
detector on the Compton Gamma Ray Observatory satellite to predict neutrino
fluences on a burst by burst basis. However, \citet{guetta} do not provide neutrino
fluences for all 73 bursts used in the triggered analysis. 

%
%

Many theoretical predictions also account for neutrino emission following
different spectral shapes both before and after the burst.  These include
precursor neutrinos coming from the GRB jet while it is still within the
progenitor \citep{choke,precursor_b} and afterglow neutrinos resulting
from interactions with the interstellar matter encountered by the relativistic
GRB jet \citep{afterglow}.  The analyses presented in this paper, however,
are optimized for the Waxman-Bahcall prompt neutrino emission spectrum only.

\section{Reconstruction and Simulation}

In both the rolling and triggered analyses, events were reconstructed with
iterative maximum likelihood reconstructions using both cascade and muon hypotheses, the latter to reject background.
The cascade hypothesis reconstruction provides a vertex, while the muon
hypothesis reconstruction provides a vertex as well as zenith and azimuth
angles. In addition to these, the triggered analysis uses a cascade hypothesis
energy reconstruction. For simulated signals we obtain a cascade vertex
resolution of about 6~m horizontally and slightly better vertically.  The cascade
energy resolution, defined as the RMS of the $\log_{10}
(E_{\mathrm{true}}/E_{\mathrm{reco}})$ distribution is approximately equal to
0.15, where $E_{\mathrm{true}}$ is the actual energy and $E_{\mathrm{reco}}$
is the reconstructed energy. For simulated downgoing muons the zenith
resolution is approximately 5$^\circ$. The down-going muon
angular resolution is worse than for other analyses because a simpler muon
reconstruction is sufficient for muon rejection. Cascade and muon reconstruction
methods are described in
\citet{kowalski:phd},\citet{taboada:phd}, \citet{ama:b10casc-pub} and 
\citet{ama:mu-reco}. Cascade reconstruction algorithms have been tested using 
artificial signals created by LEDs and lasers deployed in different locations
of the array.  These sources produce photonic signatures similar to cascades
\citep{kowalski:phd,taboada:phd}. These tests give us confidence that we understand the detector sensitivity to neutrino-induced cascades.

Both analyses used ANIS \citep{anis} for signal simulation.  All 3 neutrino
flavors were simulated with an $E^{-1}$ signal spectrum, which was then
reweighted to a broken power law.
Muon background (including multiple muons) was simulated using CORSIKA \citep{corsika}. Propagation of
muons through ice was simulated using MMC \citep{MMC} and detector response
was simulated using AMASIM \citep{amasim}. For both analyses the background is
measured experimentally (see sections \ref{section:roll} and
\ref{section:trigg}). However, background simulation was used to verify our
understanding of the detector by comparing the distribution of selection
parameters in experimental data and simulation.

\section{Rolling Analysis} 
\label{section:roll}

While satellites detect many GRBs each year, it is clear that the photonic
signatures of many bursts are missed by gamma-ray satellites. This was
especially true during the years 2001-2003, the timeframe during which the
rolling analysis was conducted, which was after BATSE ceased operations in 2000 and before
Swift launched in 2004. Rather than rely on satellite coincidence, the rolling analysis
searches for a statistical excess of events in close temporal coincidence by
sliding a time window of fixed duration over the entire data set.  Since no satellite triggers were used, this
analysis could also potentially identify neutrino signals from previously
unknown photon-dark transients and hence is not limited exclusively to GRBs.  Furthermore,
it is still an unresolved question if neutrinos arrive in coincidence with
the prompt photons or if there is a time offset. In either case, the rolling
analysis would be sensitive to GRB neutrinos.

Since BATSE results demonstrate that the distribution of GRBs is bimodal
\citep{batse}, two separate time windows were used, with durations of 1 and 100
seconds.  Although these choices do truncate the signal from some longer
bursts (assuming the neutrino burst duration is identical to the gamma burst duration), they are the most appropriate.  By studying an ensemble of real
light curves from the BATSE 4B catalog, we conclude that the gain in signal
efficiency for a small percentage of the bursts from widening the time windows
would not justify the increase in average background rate for all windows.
The numbers are kept at round values because the optimization process is not
precise enough to distinguish optimal durations to within a few percent. 

Without an external trigger, the most efficient search for a clustering of
events is conducted by starting a new window at the time of each event that
remains after cuts and counting the number of additional events in the
following 1 or 100 seconds.  

\subsection{Data Selection}

Data used in the rolling analysis come from the 2001, 2002 and 2003 AMANDA-II data
sets.  To ensure stability of the data, the austral summer periods from late
October to mid February when the South Pole station was open were omitted.
Significant work was being done on the detector and the surrounding area
at this time, which could potentially interfere with the long term stability of
the data sample during that period.  Bad files were removed from the analysis applying the same standards as AMANDA point source searches
\citep{pointsource}.  Runs less than 5000 seconds 
and files with a large number of gaps (due to unstable periods in the data) were also excluded. Deadtime percentages were 21.3\% for 2001, 15.0\% for 2002 and 15.3\% for 2003. Adjusting for deadtime, the livetimes for the datasets used in this analysis were 183.4 days for 2001, 193.8 days for 2002 and 185.2 days for 2003, yielding a total livetime of 562.4 days. 


Since there are no spatial or temporal constraints in this analysis,
background rejection is extremely important.  The first step 
is the application of a high energy filter, which cuts out events with fewer
than 160 hits\footnotemark\  or events where fewer than 72\% of optical modules
had 2 or more hits. This was followed by a process referred to as
``flare checking,'' which is designed to remove non-physical events resulting
from short-duration detector instabilities or detector malfunction
\citep{flarechecking}. \footnotetext{A ``hit'' occurs each time an optical
  module's voltage rises above a pre-set threshold, generally resulting from
  the detection of a photon.}

To further reduce the background, a loose cut was made on the variable
$N_{\mathrm{direct}}$, which is the number of hits for which there has been no
scattering of the photons in the ice. For the 2001 dataset, the exact
definition used for this cut was
$N_{\mathrm{direct}}^{\mathrm{muon}}/N_{\mathrm{hits}}$, where
$N_{\mathrm{direct}}^{\mathrm{muon}}$ is the number of direct hits using the
iterative muon fit and $N_{\mathrm{hits}}$ is the total number of hits. The
$N_{\mathrm{direct}}$ cut is useful because cascade-like events will generally
have fewer direct hits under the muon hypothesis than good muon tracks would. 
Dividing by $N_{\mathrm{hits}}$ removes the tail of high energy events which
have a large value of $N_{\mathrm{direct}}^{\mathrm{muon{}}}$ simply because
of the large number of total hits in the event.  After the 2001 data had been
analyzed, a somewhat improved cut, defined as
$(N_{\mathrm{direct}}^{\mathrm{muon}}-N_{\mathrm{direct}}^{\mathrm{cascade}})/N_{\mathrm{hits}}$,
was developed and applied to the 2002 and 2003 data sets, but was not retroactively applied to the 2001 data because this sample was previously unblinded and we did not wish to introduce trials factor penalties by altering the selection criteria. As the agreement between
data and simulation is imperfect in this variable (see Fig. 1) cutting too
close to the signal peak would introduce large systematic uncertainties.
Therefore, this variable is not included in the final cut optimization where
its position cannot be controlled, but rather used as a conservative
initial cut.


The final step in data reduction is a six variable support vector machine 
(SVM) trained with the program SVM$^{\mathrm{light}}$ \citep{svm}.  A support
vector machine uses a mathematical kernel function to find optimal cuts in a
multidimensional variable space.  The user is allowed to adjust a variable called the ``cost factor'', by which tighter or looser cuts can be obtained.  Five days of data were used from each
year as background to train the support vector machine, while ANIS simulation
was used as signal.  Cuts were finalized using only this subsample, which was
not used for the final analysis.  This was done because of the standards of
blindness applied to all AMANDA analyses. These require that all analysis criteria are decided before looking at the data in order to avoid
artificially increasing the significance of an observation through biased cut
selection.  The six variables used in the SVM are a combination of topological
cuts, which keep cascade-like signatures and reject muon signatures, and
energy-related cuts, which keep events that have properties consistent with
higher energies.  These variables are as follows:

\begin{enumerate}

\item Likelihood ratio between the muon and cascade iterative likelihood reconstructions:   This variable provides a useful means of distinguishing between events with cascade and track-like properties. This variable is shown in Figure \ref{fig:likelihood}.  As with the 5 other variables used in the support vector machine, good agreement is observed between data and background simulation. 

\item Percentage of optical modules with 8 or more hits:
   This is influenced by both the energy and type of event, as both high energy events and events producing a significant shower of particles will tend to
   produce multiple hits in each optical module. 

\item Length along the track spanned by the direct hits:  This is the length over which the direct hits
   are distributed.  This track length will naturally be shorter on average for
   the more spherically shaped cascades. 

\item $N_{\mathrm{late}}^{\mathrm{cascade}}-N_{\mathrm{late}}^{\mathrm{muon}}$:  This variable compares the number of hits which arrive more than 150 ns late relative to the fit using the cascade and muon hypotheses. 

\item $N_{\mathrm{hits}}/N_{\mathrm{OM}}$:  This variable gives the average number of
   hits per optical module with hits.  Like the percentage of modules with eight or more hits, this variable selects high
   energy cascades which produce on average more hits per module than other events. 

\item Velocity of the line fit:  The line fit is a relatively fast algorithm 
   which fits a line with velocity $v$ to each event \citep{ama:mu-reco}.  Cascade-like events will
   yield smaller velocities than muon events, which should ideally have line
   speeds close to the speed of light.
\end{enumerate}

The output of the SVM is displayed in Fig.\ref{fig:svmoutput}, showing good agreement between data and simulated background.

\subsection{Optimization}

The primary observable in the rolling analysis is $N_{\mathrm{large}}$, the largest
number of events occurring in any search window during the 3 year period.
Based on the distribution of predicted
neutrino fluences, detection of a single burst with exceptionally high
neutrino fluence is statistically more probable than detection of events from
multiple bursts.  The analysis is optimized for discovery as described in
\citet{mdf}, selecting the final cut (i.e. support vector machine cost factor) to minimize the source neutrino flux
required to produce a 5$\sigma$ observation with better than 90\% probability.
The final sensitivity, however, is only $\sim$7\% above the value obtained for
sensitivity-optimized cuts. Short and long time windows were optimized
independently.  It was assumed that background events were distributed
according to Poissonian statistics.  The data are quite consistent with
this assertion (see Fig. \ref{fig:pois}).  Background rates are not identical
over the entire year, since the downgoing muon rate varies with atmospheric
temperature. Therefore, rather than assuming a single average Poissonian
background rate, the background was characterized by using different mean
background rates for several periods during each year. 

With the chosen selection criteria, a cluster of 5 events in a 1 second window or 7 events in a 100 second window would be required for a 5$\sigma$ detection. Passing rates for the various cut stages in this analysis are shown in Table \ref{table:passingrolling}.  We now turn to a discussion of the previously mentioned triggered analysis.

\section{Triggered Analysis}
\label{section:trigg}

AMANDA-II began routine operation on Feb. 13, 2000. The last BATSE burst was
reported May 26, 2000. We have used this period of time for a coincident
search of neutrino-induced cascades and GRBs.

The Large Area Detectors (LADs) of BATSE had 4 energy channels: Channel 1: 20-50 keV, Channel 2: 50-100 keV,
Channel 3: 100-300 keV and Channel 4:~$>$~300 keV. After Feb 14, 2000, the
trigger condition for BATSE was a 5.5 sigma deviation from background on the
sum of channels 3 and 4 for three different time scales: 64 ms, 256 ms and
1024 ms. Except for one burst, GRB000213, all bursts used in this paper were
triggered as described. For GRB000213 triggering was done with channel 3 only.

Since the GRB start time, $S_{90}$, and duration, $T_{90}$, are well known,
the separation of neutrino-induced cascade signals from the down-going muon
background is simplified. We use three selection criteria based on the two
reconstruction hypotheses to discard the down-going muon background and keep
the neutrino-induced cascade signal. 

These criteria are:

\begin{enumerate}
\item Reconstructed muon zenith angle, $\theta_\mu$: This is the reconstructed
  zenith angle of the muon hypothesis. We reject events that are consistent
  with down-going muons, corresponding to $0^\circ < \theta_\mu <
  90^\circ$. For simulated cascade signals there is no correlation between
  neutrino zenith angle and the reconstructed muon zenith angle. 
\item Cascade reconstruction reduced likelihood, $L_{\mathrm{mpe}}$: This is
  the likelihood parameter (or reduced likelihood) of the multiple
  photo-electron cascade-vertex reconstruction. Smaller values correspond to
  events that match the cascade hypothesis and large values correspond to
  events that are not cascade-like. 
\item Reconstructed cascade energy, $E_{\mathrm{c}}$: This is the energy of
  the cascade hypothesis. Because the energy spectrum of the Waxman-Bahcall
  model is hard, the selection criterion $E_{\mathrm{c}}>E_{\mathrm{cut}}$ is
  good at separating signal from background. 
\end{enumerate}

A total of $\sim $7800~s per burst were studied. A period of 600 s, the 
\textit{on-time} window, centered  at the start time of the GRB, was initially 
set aside in accordance with our blind analysis procedures. The hour just
before and the hour just after the on-time window, called the
\textit{off-time} windows, are also studied.  We optimize the selection
criteria using the off-time windows and signal simulation. Thus the background
is experimentally measured. We only examined the fraction of the on-time
window corresponding to the duration of each burst. Keeping the rest of the
on-time window blind allows for other future searches, e.g. precursor
neutrinos. We use $T_{90}$ as the duration of the burst, where the time
window starts when the GRB has emitted 5\% of its total fluence and ends when
95\% have been emitted. As a precaution against possible uncertainties in the
timing of the bursts we expanded $T_{90}$ by 1~s on both sides and by the
uncertainty of the duration $U_{90}$. We will call 1~s+$T_{90}$+$U_{90}$+1~s
the \textit{signal} window. The values for $U_{90}$ were obtained from the
BATSE catalog and the typical value is 1~s.

\subsection{Data Selection}

We applied the selection criteria in two steps, a filter and the final
selection. The filter rejects down-going muons with $\theta_\mu >70^\circ$, and
keeps events that are cascade-like, $L_{\mathrm{mpe}} < 7.8$. The filter was
selected so as to maximize signal efficiency while reducing the
background. The procedure for establishing the final set of selection criteria
will be explained in section \ref{trigger_optim}. Table
\ref{table:passingtrigg} shows the passing rate of the filter.

We determined the detector stability using the off-time window experimental
data after the filter was applied. Only GRBs for which the detector is found to
be stable in the off-time windows were used for the neutrino search. 

To establish the stability of the detector, first, bad observation runs were
removed from the year 2000 data set following the same collaboration-agreed
scheme used for the rolling analysis. For GRB000508a, AMANDA fails this
test. We also checked that there are no data gaps, i.e. times the detector was off within the $T_{90}$ of the burst. For GRB000330a there
are gaps in AMANDA data. We also checked the stability of the detector by
studying the off-time windows. Two quantities were examined, the number of
events/10s that pass the filter as a function of time and the frequency
distribution of events/10s after applying the filter. Figure
\ref{fig:trigger_stability} shows the distribution of event rates around a
good burst. Visual inspection of the events/10s versus time showed a problem
with AMANDA data corresponding to burst GRB000331a. Several AMANDA strings failed to collect/report
data for periods of time on the order of 10-100~s. For this reason, GRB000331a was
excluded from this analysis. We have also looked at the plots of time
difference between events to check for possible detector problems. No new
problems were found. Finally, we
also exclude GRB000217a and GRB000225 from the list of bursts because AMANDA
was not operational for these two bursts.

After all these criteria are used we find 73 BATSE bursts for which the
detector is behaving stably. Of these bursts, 53 are long bursts 
($T_{90}>2$~s) and 20 are short bursts ($T_{90}<2$~s). In the BATSE catalog
\citep{batse} $T_{90}$ were not available for 13 bursts. The lack of
$T_{90}$ may be caused by gaps in the BATSE data not being treated properly by 
automatic procedures. In this case the light curves for the bursts without
$T_{90}$ were obtained from BATSE's web  page. The comments in the web page
were also studied. Based on visual inspection of the light curves and the
comments, conservative, i.e. large, values for burst duration were chosen. For 12 of
the bursts with missing $T_{90}$ we examined the light curves for the combined
channels 1-4. For burst GRB000517 we used the light curve for the combined
channels 1-3 since channel 4 was missing. Table \ref{table:bursts} summarizes
the characteristics of the 78 bursts (73 used in this analysis) reported by
BATSE between February 13 and May 26, 2000.

\subsection{Optimization}
\label{trigger_optim}

The selection criteria were optimized on the off-time windows for discovery in
a procedure similar to that of the rolling analysis but with the difference
that we optimize two selection criteria simultaneously. The final selection
criteria are $L_{\mathrm{mpe}}<6.9$ and $E_{\mathrm{c}}>40$~TeV. Figure
\ref{fig:filter} shows the $L_{\mathrm{mpe}}$ and $E_{\mathrm{c}}$
distribution after the filter has been applied for the data in the
\textit{signal} window, along with simulated background and simulated neutrino
signal.  After all selection criteria are applied, one event remains in the 73
burst combined off-time windows. This is equivalent to an expected background
of $n_{\mathrm{b}}=0.0054^{+0.013}_{-0.005}$~(stat) in the 73 burst combined
\textit{on-time} window. Passing rates for the various cut stages in this
analysis are shown in Table \ref{table:passingtrigg}.  Three or more on-time, on-source events would be required for a 5$\sigma$ detection.

The total \textit{signal} window is 2851.44~s corresponding to 2591.61~s for
$T_{90}$, 113.83~s for the sum of the uncertainty on $T_{90}$ and 146~s for
the padding of the on-time window. The total \textit{off-time} window is
529329~s. For the specific set of runs used in this search the AMANDA-II
dead-time is 17.8\%.

\section{Systematic Uncertainties} \label{section:sysun}

Multiple effects have to be considered when estimating the systematic
uncertainties: properties of the ice, detector effects, neutrino-matter
cross-sections, etc. We have used signal simulations to estimate the
uncertainties and artificial light sources to verify that the
detector is sensitive to cascade-like signals \citep{kowalski:phd,taboada:phd}.

The actual optical properties of ice at the South Pole are known
with a reasonably high degree of precision \citep{theicepaper}, but
this knowledge is not fully incorporated into the signal simulation software
that was available for this paper. The IceCube collaboration is working on
improved simulation software so that better optical ice models are available for future analyses.

To estimate the systematic uncertainty due to the optical properties of the ice we
have performed signal simulations supposing a Waxman-Bahcall spectrum using
the most and least transparent ice that has been measured at AMANDA
depths. In the triggered analysis we find 30\% more signal events than with
average optical properties for the clearest ice and  we find 65\% fewer events
than with average optical properties for the least transparent ice. In the
rolling analysis we find 50\% more signal events in the clearest ice and 50\%
fewer events in the least transparent ice. 
It should be noted that these
systematic uncertainties are not RMS ranges,
instead they are extreme values. We will suppose that systematic
uncertainties have a flat distribution between the extrema found. The
equivalent RMS values are $^{+9\%}_{-19\%}$ for the triggered analysis
and $\pm 14$\% for the rolling analysis. The systematic uncertainties due to
ice properties in this paper are larger than in our previous publications on
neutrino-induced cascades \citep{ama:b10casc-pub,ama2:cascades}. This is
because for the previous publications an $E^{-2}$ spectrum was assumed. For
hard spectra such as Waxman-Bahcall (see Equation \ref{eqn:wb}), the
uncertainty due to optical properties of ice is larger. Additionally, we use
different selection criteria.

We followed a similar procedure for estimating the effect of the uncertainty
in the absolute efficiency of the optical modules. A 10\%
uncertainty in the absolute efficiency results in a change of 
3\% in the number of signal events in the triggered analysis and a 5\% change
in the rolling analysis. Similarly, a 5\% uncertainty in the
neutrino-matter cross section \citep{gandhi} results in a 4\% change in
the number of signal events. Other effects like OM pre-pulsing
\citep{ama:b10casc-pub}, electronic crosstalk, and differences between data
and simulation  make negligible contributions to the systematic uncertainties.

In the case of the rolling analysis, there is also a
$\pm$20\% percent uncertainty in the final limit resulting from the uncertainty
in the burst-by-burst spread of neutrino fluxes. This uncertainty results from
several factors, primarily variations in the distribution of events depending
on what model parameterizations are used and uncertainty in the fit applied to
the data.  This procedure is explained in more detail in section
\ref{section:rollingresults}.

We thus find that the simulation of optical properties of ice is the
single most important contribution to the systematic uncertainties. Adding in
quadrature the signal systematic uncertainties results in a global signal
uncertainty of $^{+31\%}_{-65\%}$ for the triggered analysis and
$\pm 54\%$ for the rolling analysis.

\section{Results}

For both the rolling and triggered analysis we do not find evidence of
neutrino-induced cascades from gamma-ray bursts. We derive limits on the total diffuse neutrino flux due to all GRBs using the
\cite{feldman-cousins} unified procedure. We include systematic uncertainties
following \citet{sysun} and \citet{hill}. Our limits depend on the modeling of
the distribution with redshift of gamma-ray bursts \citep{redshift}. For the
triggered analysis the models use burst distributions that follow the
experimental selection effects of BATSE. The rolling analysis is not
constrained by these selection effects and thus long duration bursts should be
modeled as following the star formation rate. In practice, however, we use the
same distribution for both analyses because the difference between the two
options is extremely small. This is probably because only bursts with
relatively high fluence contribute significantly to the neutrino flux.

We present Model Rejection Factors\footnotemark, MRF, \citep{mrf} for
\footnotetext{ The Model Rejection Factor is the multiplicative factor by
  which a predicted flux would need to be scaled in order to be ruled out by
  an analysis at a 90\% confidence level.} 
\citet{wb}, \citet{supranova}, \citet{choke} and \citet{murase} model A. For
the Waxman-Bahcall model we assume 1:1:1 flavor flux ratio, p$\gamma$ 
neutrino generation, 666 bursts per year and a flux normalization\footnotemark
of $A_{\nu_e+\nu_\mu+\nu_\tau} =
1.3\times$10$^{-8}$~GeV~cm$^{-2}$~s$^{-1}$~sr$^{-1}$. We ignore the transition 
\footnotetext{Note that it is also possible to base the normalization on the
  average photon fluence (as opposed to ultra high energy cosmic rays) of
  $F_{\gamma} \sim 6 \times 10^{-6}$ erg/cm$^2$ and 666 bursts per year as
  observed by BATSE. This results in a flux normalization of
  $A_{\nu_e+\nu_\mu+\nu_\tau} = 
  2.3\times$10$^{-9}$~GeV~cm$^{-2}$~s$^{-1}$~sr$^{-1}$ including all flavors 
  and  oscillations. This normalization takes into account the selection
  effects of BATSE.}  
from 1:1:1 flux ratio to 1:1.8:1.8 with increasing energy, which would change the limits by
$\sim$10\% in both analyses. For the \citet{supranova} supranova model we
assume 445 bursts per year (or 2/3 of 666, the fraction of long duration
bursts), pp neutrino generation below 2~PeV and p-$\gamma$ above this
energy. It should be noted that this supranova model is not well supported by
observational data because it assumes a delay of $\sim$1~week to several
months between the supernova and the GRB. Observations of supernovae
associated with gamma-ray bursts, e.g. GRB060218, have placed limits to this
delay to be as small as a few hours \citep{GRB060218}. Model A of
\citet{murase} assumes that the GRB rate is tied to star formation rate.  We have also been provided
\citep{murase-priv} with the flux for the same model but for bursts following
the redshift distribution of long duration BATSE-like bursts. In both analyses we use the latter distribution, which corresponds to a rate of 445 long-duration bursts per year.  In practice, the difference in the predicted neutrino spectra in these two cases is very small. The model parameters
used include a beamed energy per burst of 2$\times$10$^{51}$~ergs and the baryon
loading factor is taken to be 100, a value which assumes GRBs are the primary
source of cosmic rays. It should be noted that, since \citet{murase} Model A is
available for both electron and muon neutrino fluxes at the source, for this model
these fluxes are used to calculate the flavor flux ratio at Earth taking into account full neutrino mixing. Because the electron and muon flux spectra are different, the flavor flux ratio at Earth is not strictly 1:1:1 for this model, but rather varies as a function of energy.


Limits for models that are not presented here can be tested by calculating:

\begin{equation}
N_{\mathrm{expected}} = T \int dE_{\nu} d\Omega \phi(E_\nu) A_{\mathrm
  {eff}}(E_\nu,\theta_\nu),  
\end{equation}
where $T$ is the exposure time, $\phi$ is the neutrino flux at the Earth's
surface according to the model, $A_{\mathrm{eff}}$ is the effective area and
$N_{\mathrm{events}}$ is the number of events predicted by the model. Given an
expected number of events and the 90\% c.l. upper signal event limit,
$N_{90}$, the MRF for the model to be tested is:

\begin{equation}
\mathrm{MRF} = \frac{N_{90}}{N_{\mathrm{expected}}}.
\end{equation}

Figures \ref{fig:effarearoll} and \ref{fig:effareatrig} show the neutrino
effective area of AMANDA after all selection criteria for the rolling and the
triggered analyses respectively have been applied.

\subsection{Rolling Analysis}
\label{section:rollingresults}

Upon unblinding the rolling analysis, the maximum number of events observed in
any bin for the 1 second search was 2, while the maximum in any bin in the 100
second search was 3.  These were the most likely outcomes of the analysis
assuming no signal was present (with probabilities 70.2\% and 75.4\%,
respectively,  based on computer simulation). Further, the number of doublets
and triplets, i.e. 2 or 3 events in a single time window, was very consistent
with predictions assuming Poissonian statistics.  The number of doublets in
the 1 second search was 311 on an expected background of 310$\pm$20.  The
number of doublets in the 100 second search was 1000 on an expected background
of 1020$\pm$30 and the number of triplets was 20 on an expected background of
22$\pm$5.

Because this analysis looks for a cluster of
temporally correlated events, it is not just the overall neutrino flux that
determines the level at which we can observe a neutrino signal, but also the
way that the neutrino flux is divided among discrete bursts. For example,
it is statistically much more probable to obtain a cluster of several events
from one very strong, nearby burst than from 100 bursts occurring at different
times, even if the net neutrino fluxes at Earth for the two scenarios are
equivalent. It is therefore necessary in this case to make an assumption about
the relative distribution of neutrino events among all GRBs. Thus, the MRF for each model tested is
determined using a signal simulation which varies the average expected
neutrino flux by a random factor for each burst. These factors are weighted
according to a gaussian fit to the distribution of predicted event rates for individual
GRBs from the BATSE catalog, which were obtained from \citet{guetta}. This
accounts for several factors affecting neutrino flux, including distance from
Earth and electromagnetic fluence. The majority of bursts therefore have a
signal flux near the average rate while a few have either much higher or lower
fluxes.  The total year-long flux is thus divided into a number of unequal discrete bursts, with the number of bursts per year determined from the burst rate observed by  BATSE.

The MRF for the Waxman-Bahcall method is 120 (100 without systematics), with
90\% of events in the 70 TeV to 8 PeV energy range. For this model, 1/3 of the bursts were assumed to be short (applied to both time windows) and 2/3 assumed to be long (applied to the 100 second time window only), with corrections made for the lower average fluence from short bursts relative to long bursts.  The MRF relative to the \citet{supranova} supranova model is 27, while relative to the Murase-Nagataki Model A flux, the MRF is 95.  Since these models pertain only to long bursts, only the 100 second window was used for these models and the number of bursts per year was assumed to be 445.

One possible additional class of sources without direct photon signatures is
choked bursts, which would emit precursor neutrinos like a conventional GRB,
but have no $\gamma$-ray emission or prompt neutrinos because the
fireball 
never escapes from the interior of the stellar progenitor \citep{choke}.  The
rolling analysis cuts are not optimized for the energy spectrum predicted for
choked bursts, which peaks at a few TeV rather than $\sim$100 TeV. The MRF
calculated for this model is 72, 
assuming a choked burst rate 100 times greater than the rate of conventional
GRBs (tied to the rate of type II supernovae) and assuming the progenitor to
have an external hydrogen envelope \citep{precursor_b}. Figure
\ref{fig:limits} and Table \ref{table:limits} summarize the limits presented here.

\subsection{Triggered Analysis}

After applying all selection criteria, for a simulated Waxman-Bahcall spectrum,
we expect 0.03 events from 73 bursts. The final events sample is composed 55\%
by $\nu_e$ and $\bar{\nu}_e$, 7\% by $\nu_\mu$ and $\bar{\nu}_\mu$ and 38\% by
$\nu_\tau$ and $\bar{\nu}_\tau$. The central 90\% of the events from the
Waxman-Bahcall flux are between 70 TeV and 8 PeV. Taking into account that the
ratio of signal to off-time windows is 5.387$\times$10$^{-3}$, we expected a
background of 0.0054$^{+0.013}_{-0.005}$.

After examination of the \textit{signal} windows, no events are found in the
combined 73 signal windows, so we find no evidence for neutrino
induced-cascades in coincidence with GRBs reported by BATSE from February 13,
2000 to May 26, 2000. The signal event upper limit $N_{90}$ is 3.5 (2.4
without systematics).

In order to determine what fraction of the total year-long isotropic neutrino flux comes from the bursts included in our sample, we simply divide the number of bursts studied by the expected total number of relevant bursts occurring per year.  For the Waxman-Bahcall model, we have included both long duration GRBs and
short duration GRBs, because the original model does not distinguish between
the two classes.  Thus, we assume our 73 burst sample contains 73/666, or 11\%, of the year's total neutrino flux.  The supranova and Murase-Nagataki models, however, apply only to long bursts.  Since there are 53 long bursts in our sample and an expected rate of 445 long bursts per year, 12\% of the total long burst neutrino flux is assumed to be contained in our burst sample.

The MRF for the Waxman-Bahcall model
is 110 (78 without systematics).  For the supranova model the expected signal after applying all selection critera is 0.067  and the MRF, corrected by systematic uncertainties, is 25. For
Murase-Nagataki Model A we expect a signal of 0.0038 events.  This signal expectation corresponds to a MRF of 94.  


\section{Conclusions}

We have performed two searches for neutrino-induced cascades with
AMANDA-II. The 
triggered analysis searched for neutrinos in coincidence with 73 gamma-ray
bursts reported by BATSE in 2000. The rolling analysis searched for a
statistical excess of cascade-like events in time rolling windows of 1 and 100
s for the years 2001, 2002 and 2003. No evidence for neutrino-induced cascades
from gamma-ray bursts is found. We present MRFs for the 
Waxman-Bahcall model, the supranova model, a choked-burst model and Murase and
Nagataki Model A. For the Waxman-Bahcall model the MRF is 110 from the
triggered analysis and 120 from the rolling analysis.  At 1~PeV the triggered
analysis limit is:
\begin{equation}
E^2 \frac{d\Phi}{dE} \leq
1.5 \times 10^{-6}\; \mathrm{GeV cm^{-2} s^{-1} sr^{-1}}, 
\end{equation}
and the rolling analysis limit is:
\begin{equation}
E^2 \frac{d\Phi}{dE} \leq
1.6\times10^{-6}\; \mathrm{GeV cm^{-2} s^{-1} sr^{-1}}.  
\end{equation}

Although there are advantages to the search methods discussed in this paper, 
our limits are not as constrictive as the muon neutrino limit, which lies at
1.7$\times$10$^{-8}$~GeV~cm$^{-2}$~s$^{-1}$~sr$^{-1}$ for the
Waxman-Bahcall spectrum at 1 PeV \citep{GRB-muon}.  This value is for a single neutrino flavor
only and should therefore be multiplied by a factor of $\sim$3 to obtain a
more direct comparison to cascade all-flavor limits.

For the triggered analysis this difference is in large part due to the fact
that the neutrino-induced muon search uses a much higher number of
approximately 400 bursts reported between 1997 and 2003. Because the triggered
analysis has a very low background rate the sensitivity should grow linearly
with the number of bursts studied. Given the same set of bursts, the
sensitivity of the triggered analysis is only a factor $\sim$4 worse than that
of the neutrino-induced muon search. But, unlike the triggered up-going muon
search, the triggered cascade analysis is sensitive to gamma-ray bursts in
both the Southern and Northern Hemisphere. This can potentially double the 
sensitivity. In the case of the rolling analysis, the lack of spatial and
temporal constraints results in a reduced per-burst sensitivity relative to
triggered analyses, yet allows it to sample from a larger group of
transients. This analysis therefore has the potential to detect sources missed
by other methods. It thus serves as a useful complement to triggered GRB
searches, especially during periods without large satellite experiments
dedicated to GRB study. It should be noted that AMANDA searches for diffuse
fluxes of extraterrestrial neutrinos using cascades
\citep{ama2:cascades,diffuse} can also be used to establish limits on
neutrino emission by GRBs. But given the same exposure the analyses presented
here have better sensitivity because time correlations significantly reduce
the background.

Future searches with the AMANDA and IceCube detectors may include bursts 
reported by Swift, GLAST and other IPN satellites. The capabilities of IceCube are particularly
promising. Preliminary studies indicate that a triggered search for 300-500
bursts with IceCube would suffice to set limits at levels lower than predicted
by Waxman-Bahcall or would find evidence of the existence of neutrinos in
coincidence with GRBs with better than 5$\sigma$ confidence. Also, bursts that
are particularly bright and close may result in signals that are strong enough
to provide an unequivocal discovery from a single burst \citep{nus-030329}. If
such a burst were to occur in the southern sky, only the cascade channel would
be available to study this burst. 

\acknowledgments

We are grateful for data provided and comments made by Dr. K. Murase. We
acknowledge the support from the following agencies: National Science
Foundation-Office of Polar Program, National Science Foundation-Physics
Division, University of Wisconsin Alumni Research Foundation, Department of
Energy, and National Energy Research Scientific Computing Center (supported by
the Office of Energy Research of the Department of Energy), the NSF-supported
TeraGrid system at the San Diego Supercomputer Center (SDSC), and the National
Center for Supercomputing Applications (NCSA); Swedish Research Council,
Swedish Polar Research Secretariat, and Knut and Alice Wallenberg Foundation,
Sweden; German Ministry for Education and Research, Deutsche
Forschungsgemeinschaft (DFG), Germany; Fund for Scientific Research
(FNRS-FWO), Flanders Institute to encourage scientific and technological
research in industry (IWT), Belgian Federal Office for Scientific, Technical
and Cultural affairs (OSTC); the Netherlands Organisation for Scientific
Research (NWO); M. Ribordy acknowledges the support of the SNF (Switzerland);
J. D. Zornoza acknowledges the Marie Curie OIF Program (contract 007921).

\clearpage
\begin{deluxetable}{lrr}
\tabletypesize{\small}
\tablewidth{0pt} 
\tablecaption{\label{table:passingrolling}Passing rates for experimental data and simulated Waxman-Bahcall spectrum, $\nu_e + \bar{\nu}_e$.}
\tablehead{
\colhead{} & \colhead{Exp Data} & \colhead{$\nu_e + \bar{\nu}_e$}}
\startdata
Initial & 100\% & 100\% \\
Filter & 0.80\% & 62\% \\
$N_{dir}$ cut & 0.10\% & 62\% \\
SVM short window search & 0.0027\% & 58\% \\
SVM long window search & 0.00040\% & 43\% \\
\enddata
\end{deluxetable}

\clearpage

\begin{deluxetable}{lrr}
\tabletypesize{\small}
\tablewidth{0pt} 
\tablecaption{\label{table:passingtrigg}Simulated $\nu_e + \bar{\nu}_e$
  passing rates following a Waxman-Bahcall spectrum and \textit{off-time} window passing
  rates for the triggered analysis.}
\tablehead{
\colhead{} & \colhead{Off-time} & \colhead{$\nu_e + \bar{\nu}_e$}}
\startdata
Initial & 100\% & 100\% \\
Filter & 0.91\% & 67\% \\
$L_{\mathrm{mpe}}<6.9$ & 0.05\%& 35\%\\
$E_{\mathrm{c}} > 40$~TeV & 4$\times 10^{-6}$\% & 25\%\\
\enddata
\end{deluxetable}

\clearpage

\begin{deluxetable}{clrrr}
\tabletypesize{\small}
\tablewidth{0pt} 
\tablecaption{\label{table:bursts}List of Bursts used for the triggered
  analysis.}
\tablehead{
\colhead{BATSE ID} & \colhead{Burst} & \colhead{$T_{90} (s)$} & \colhead{RA (J2000) (deg)} & \colhead{Dec (J2000) (deg)}
}
\startdata
7988 &	GRB000213 &	0.41 &	4.80 &	225.14 \\
7989 &	GRB000217a &	30.57 &	36.51 &	126.25 \\
7990 &	GRB000217b & n/a \tablenotemark{d} & -56.97 & 337.12 \tablenotemark{c}\\
7991 &	GRB000219 &	1.00 &	84.14 &	116.37 \tablenotemark{a}\\
7992 &	GRB000220 &	2.45 &	65.95 &	129.86 \\
7994 &	GRB000221 &	26.18 &	77.70 &	136.20 \\
7995 &	GRB000222 &	0.61 &	60.60 &	141.82 \\
7997 &	GRB000225 &	16.70 &	0.53 &	215.99 \tablenotemark{c}\\
7998 &	GRB000226a &	10.24 &	29.82 &	197.28 \\
7999 &	GRB000226b &	0.53 &	16.89 &	74.58 \\
8001 &	GRB000227 &	75.14 &	-7.49 &	184.37 \\
8002 &	GRB000228 &	15.00 &	65.16 &	99.50 \tablenotemark{a}\\
8004 &	GRB000229 &	32.51 &	47.87 &	81.33 \\
8005 &	GRB000301 &	25.00 &	72.68 &	120.17 \tablenotemark{a}\\
8008 &	GRB000302a &	22.66 &	54.28 &	147.47 \\
8009 &	GRB000302b &	14.34 &	30.66 &	196.18 \\
8012 &	GRB000303 &	17.66 &	62.05 &	91.46 \\
8018 &	GRB000306a &	0.13 &	-10.17 &	206.83 \\
8019 &	GRB000306b &	51.20 &	40.92 &	68.39 \\
8022 &	GRB000307 &	22.53 &	6.80 &	200.18 \\
8026 &	GRB000310a &	327.30 & -10.86 & 234.59 \\
8027 &	GRB000310b &	1.54 &	-1.46 &	106.10 \\
8030 &	GRB000312a &	23.87 &	37.92 &	83.64 \\
8031 &	GRB000312b &	45.00 &	11.04 &	200.09 \tablenotemark{a}\\
8033 &	GRB000313a &	0.13 &	-19.37 & 343.91 \tablenotemark{a}\\
8035 &	GRB000313b &	0.77 &	10.25 &	319.57 \\
8036 &	GRB000314 &	110.85 & 50.66 & 167.77 \\
8039 &	GRB000317 &	83.52 &	32.66 &	136.70 \\
8041 &	GRB000319 &	0.08 &	-13.86 & 275.00 \\
8045 &	GRB000320 &	44.16 &	4.44 & 199.27 \\
8047 &	GRB000321 &	0.89 &	36.39 &	153.04 \\
8049 &	GRB000323 &	72.45 &	48.08 &	126.91 \\
8050 &	GRB000324 &	3.90 &	-24.04 & 319.19 \\
8053 &	GRB000326a &	1.92 &	-26.36 & 24.96 \\
8054 &	GRB000326b &	21.25 &	-63.47 & 330.45 \\
8056 &	GRB000330a &	26.00 &	32.00 &	74.84 \tablenotemark{c}\\
8057 &	GRB000330b &	0.40 &	39.26 &	110.80 \tablenotemark{a}\\
8058 &	GRB000331a &	25.00 &	-15.02 & 271.73 \tablenotemark{c}\\
8059 &	GRB000331b &	78.66 &	-46.29 & 290.09 \\
8061 &	GRB000331c &	26.94 &	59.77 &	132.44 \\
8062 &	GRB000401 &	133.44 & 80.60 & 112.87 \\
8063 &	GRB000402 &	106.62 & 6.65 &	78.59 \\
8064 &	GRB000403 &	148.22 & 24.69 & 166.48 \\
8066 &	GRB000407 &	28.93 &	-70.06 & 291.50 \\
8068 &	GRB000408a &	0.62 &	-71.85 & 319.61 \\
8069 &	GRB000408b &	4.78 &	67.22 &	146.61 \\
8071 &	GRB000409 &	41.34 &	80.82 &	112.91 \\
8072 &	GRB000410 &	0.35 &	-12.48 & 327.83 \\
8073 &	GRB000412 &	33.02 &	-59.78 & 307.21 \\
8074 &	GRB000415a &	11.00 &	68.27 &	132.37 \tablenotemark{a}\\
8075 &	GRB000415b &	20.80 &	69.42 &	144.65 \\
8076 &	GRB000415c &	0.22 &	-29.98 & 309.64 \\
8077 &	GRB000417 &	1.66 &	2.93 & 357.46 \\
8079 &	GRB000418 &	2.29 &	76.15 &	135.19 \\
8080 &	GRB000420a &	140.00 & -44.66 & 267.84 \tablenotemark{a}\\
8081 &	GRB000420b &	46.00 &	-14.59 & 238.81 \tablenotemark{a}\\
8082 &	GRB000420c &	10.11 &	-63.01 & 332.47 \\
8084 &	GRB000421 &	82.18 &	16.98 &	240.68 \\
8085 &	GRB000424a &	3.58 &	71.80 &	107.62 \\
8086 &	GRB000424b &	18.43 &	53.98 &	162.56 \\
8087 &	GRB000429 &	164.35 & -4.81 & 216.02 \\
8089 &	GRB000502 &	0.12 & -46.68 &	339.87 \\
8097 &	GRB000508a &	1.00 &	3.78 & 326.62 \tablenotemark{c}\\
8098 &	GRB000508b &	136.19 & -20.38 & 0.51 \\
8099 &	GRB000508c &	15.49 &	2.39 & 204.79 \\
8100 &	GRB000509 &	20.00 &	-39.27 & 358.61 \tablenotemark{a}\\
8101 &	GRB000511a &	115.01 & -36.11 & 8.02 \\
8102 &	GRB000511b &	38.98 &	-8.70 &	30.83 \\
8104 &	GRB000513a &	0.38 &	-45.11 & 350.24 \\
8105 &	GRB000513b &	11.33 &	-12.01 & 260.19 \\
8109 &	GRB000517 &	51.00 &	76.74 &	137.86 \tablenotemark{a}\\
8110 &	GRB000518 &	10.30 &	53.91 &	153.22 \\
8111 &	GRB000519 &	14.59 &	3.33 &	78.40 \\
8112 &	GRB000520 &	14.98 &	-0.31 &	5.64 \\
8113 &	GRB000521 &	2.00 &	-6.25 &	104.25 \tablenotemark{a}\tablenotemark{b}\\
8116 &	GRB000524 &	49.98 &	-41.36 & 252.93 \\
8120 &	GRB000525 &	1.41 &	-39.44 & 355.92 \\
8121 &	GRB000526 &	36.86 &	-10.32 & 353.05 \\
\enddata
\tablenotetext{a}{Duration selected by visual inspection of the light curves}
\tablenotetext{b}{We classify this burst as short}
\tablenotetext{c}{Burst not used for triggered analysis} 
\tablenotetext{d}{No $T_{90}$ in catalog}
\end{deluxetable}

\clearpage

\begin{deluxetable}{cccc}
\tabletypesize{\small}
\tablewidth{0pt} 
\tablecaption{\label{table:limits}Model Rejection Factors.}
\tablehead{
\colhead{Model} & \colhead{Triggered Analysis} & \colhead{Rolling 
    Analysis} & \colhead{Energy Range (90\% of events)}
}
\startdata
Waxman-Bahcall & 110 & 120 & 70 TeV to 8 PeV\\
Razzaque et al.& 25  &  27 & 50 TeV to 7 PeV\\
Murase-Nagataki (model A) & 94 &  95 & 100 TeV to 10 PeV \\
Choked Bursts & n/a &  72 & 8 TeV to 61 TeV \\
\enddata
\end{deluxetable}

\clearpage

\begin{figure}
\begin{center}
\plotone{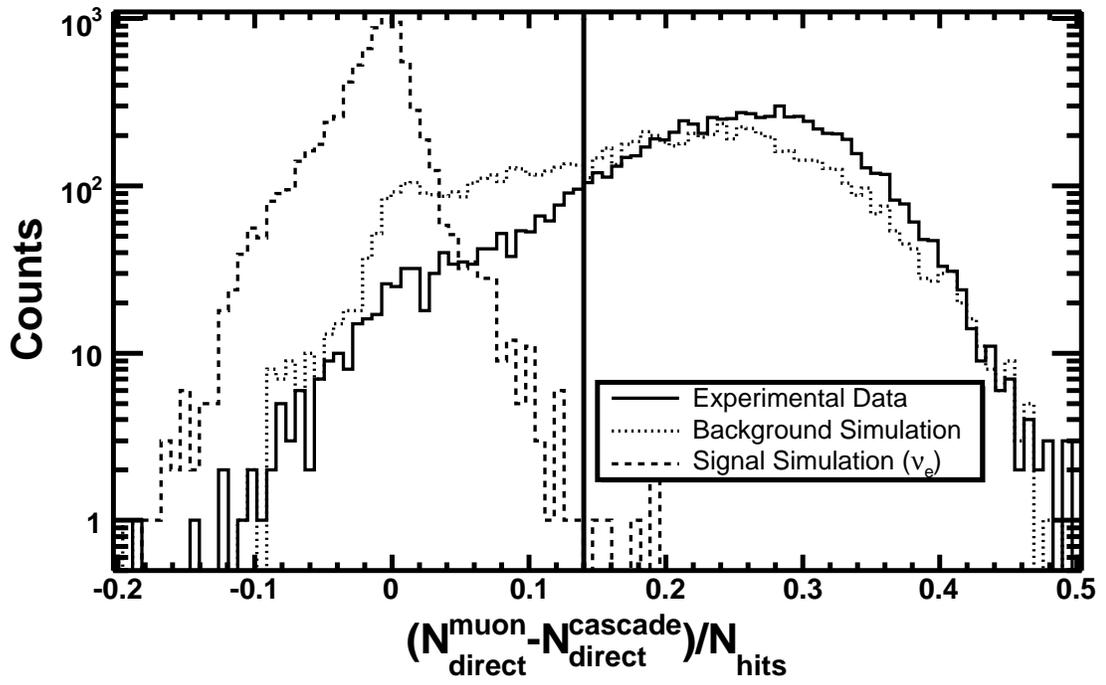}
\caption{\label{fig:ndird}The cut variable
  (N$_{\mathrm{dir}}^{\mathrm{muon}}$-N$_{\mathrm{dir}}^{\mathrm{cascade}}$)/N$_{\mathrm{hits}}$.
  Values above 0.14 are removed.  N$_{\mathrm{dir}}$ is the number of hits for which there has been no scattering of the photons in the ice.}
\end{center}
\end{figure}

\begin{figure}
\begin{center}
\plotone{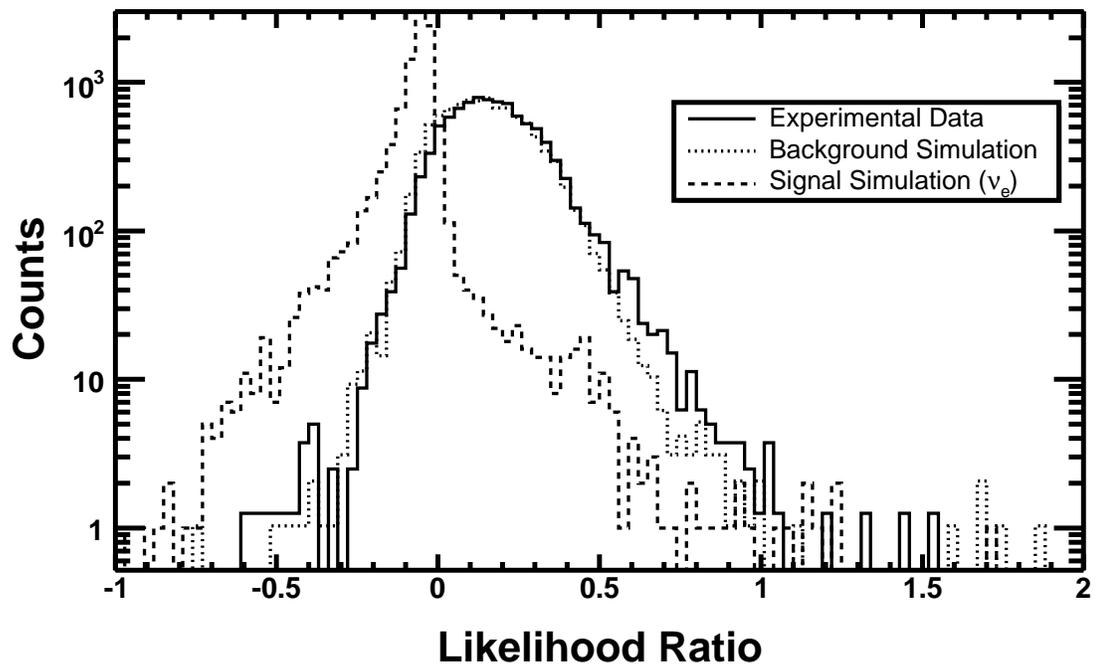}
\caption{\label{fig:likelihood}The likelihood ratio compares the likelihood of a given event being a muon to the likelihood of it being a cascade.  This variable is shown as a representative example of the six variables used in the support vector machine cut.}
\end{center}
\end{figure}

\begin{figure}
\begin{center}
\plottwo{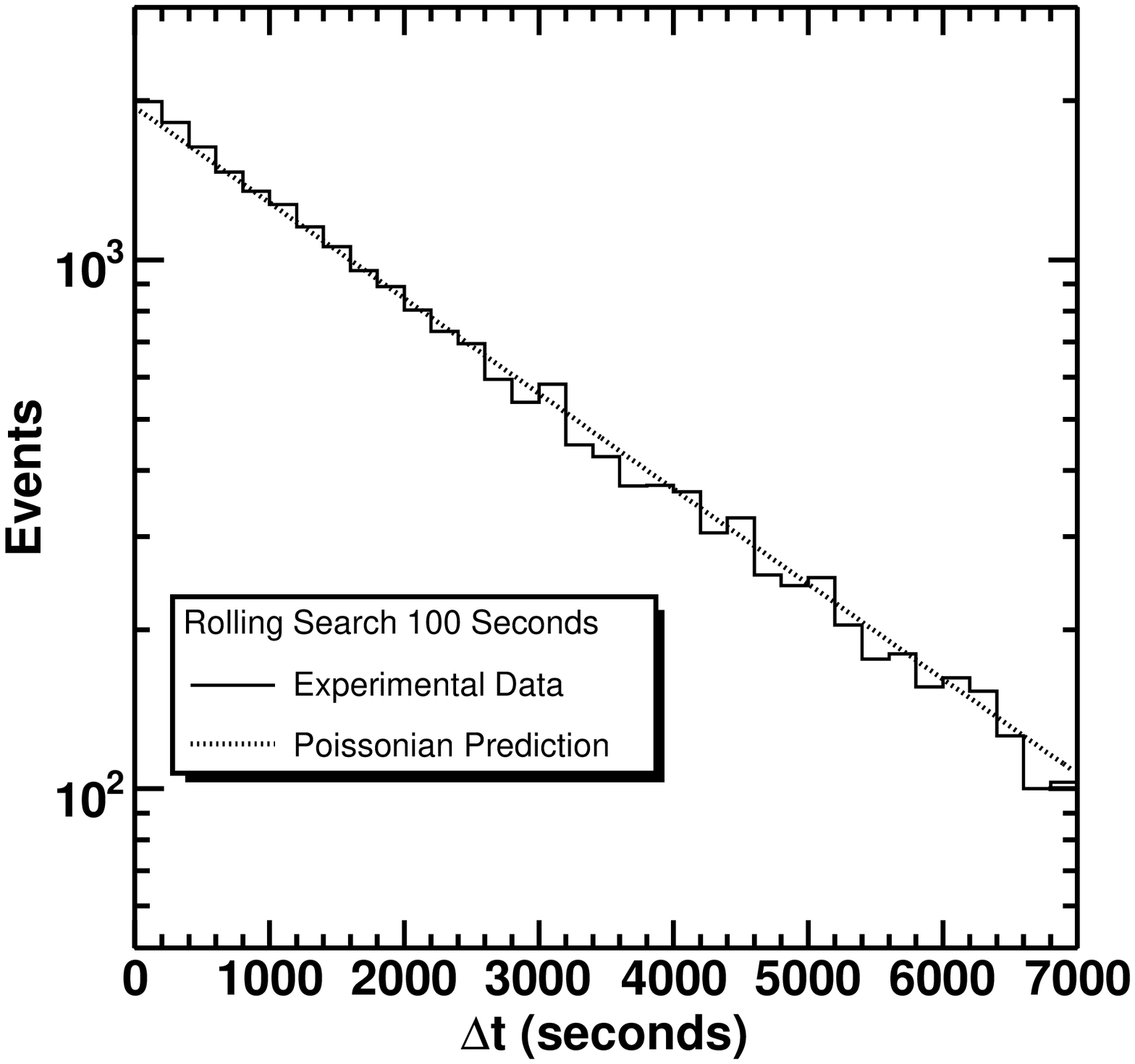}{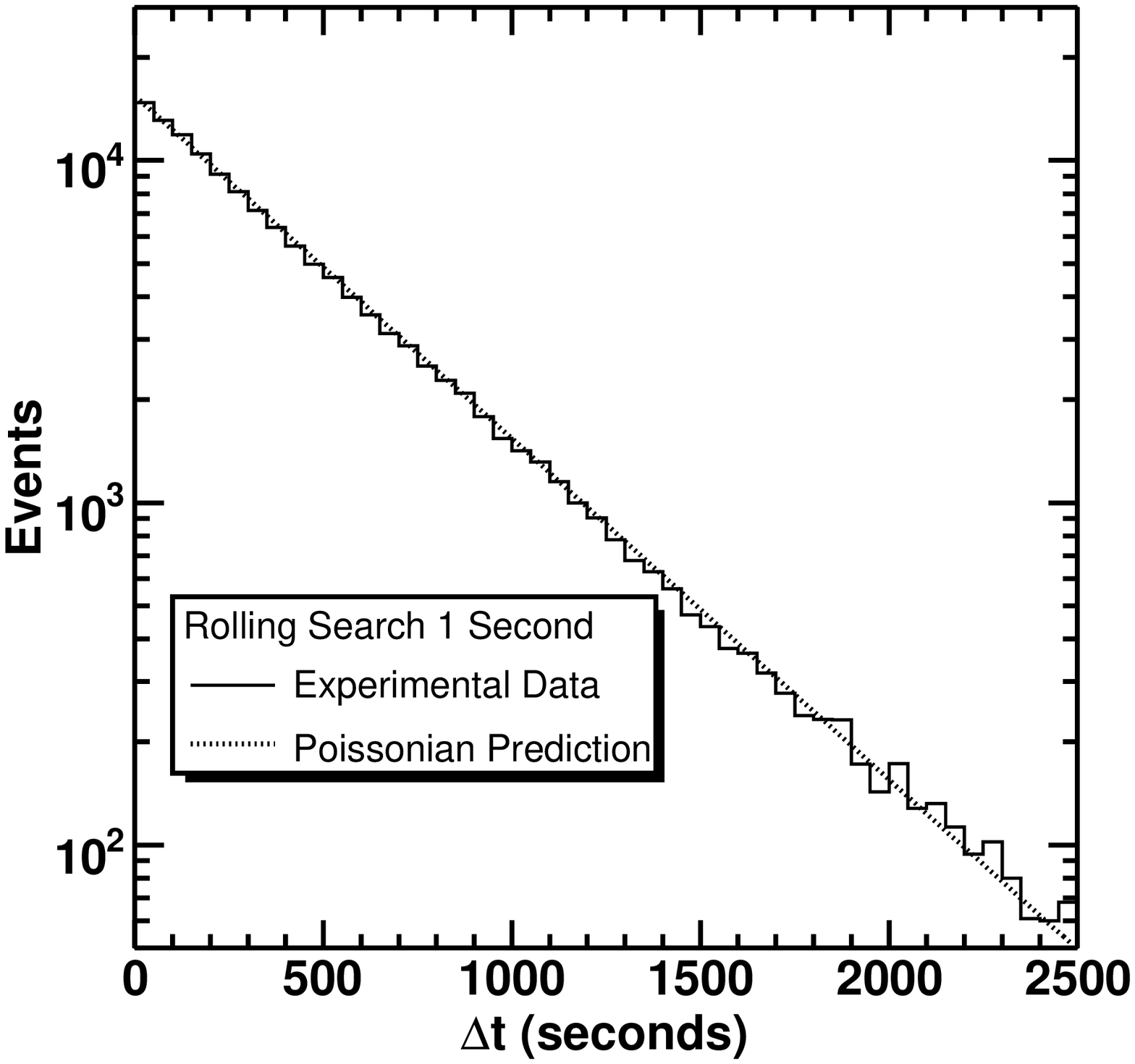}
\caption{\label{fig:pois} Time difference $\Delta$t between surviving events for both the 100 second (left)and 1 second (right) searches.  The solid line shows experimental data for all three years in which the analysis was conducted.  The dotted line
  shows the theoretical prediction, modeling the background with a Poisson
  distribution and dividing each year into 5 periods with unique Poissonian
  average rates. Because the two time windows were optimized independently,
  these curves correspond to different average event rates: 1 event per 2404
  seconds for the long window search (left) and 1 event per 427 seconds for
  the short window search (right).}
\end{center}
\end{figure}

\begin{figure}
\begin{center}
\plotone{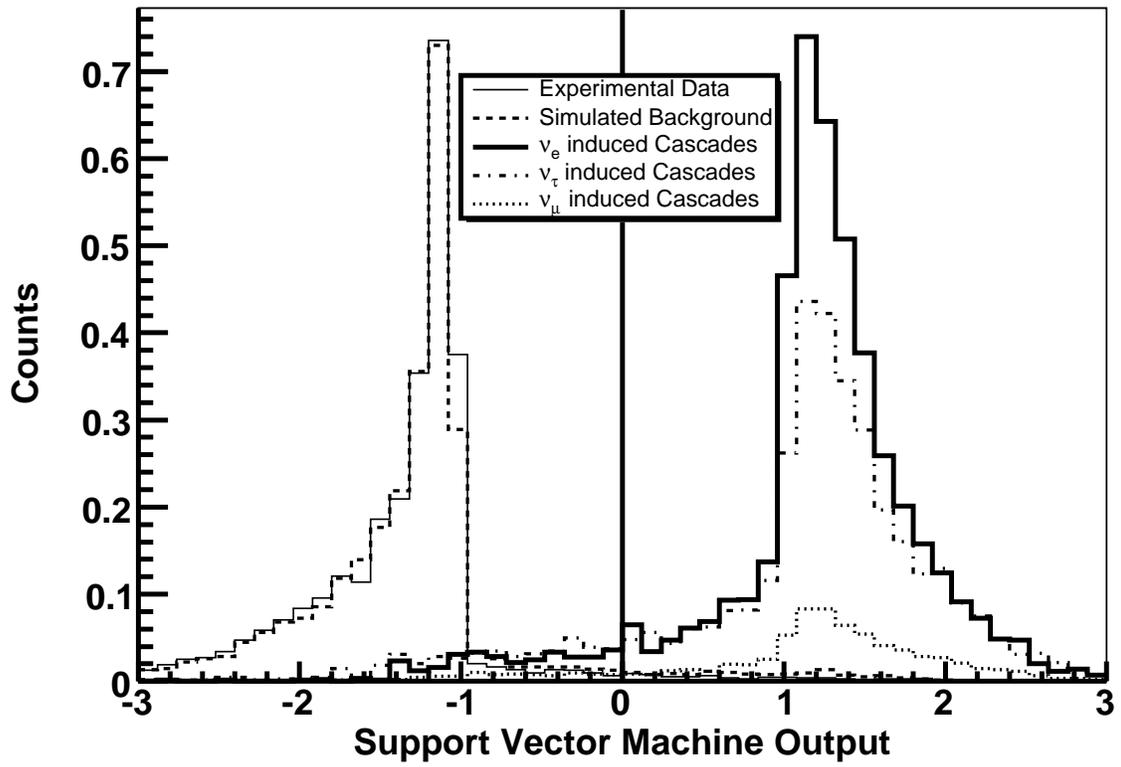}
\caption{\label{fig:svmoutput}Support Vector Machine output for experimental data, simulated background and simulated signal resulting from
  the three neutrino flavors.  Values above zero are considered signal, while
  those below zero are considered background and rejected. Muon neutrino
  signal simulation corresponds to neutral current interactions .} 
\end{center}
\end{figure}

\begin{figure}
\begin{center}
\plottwo{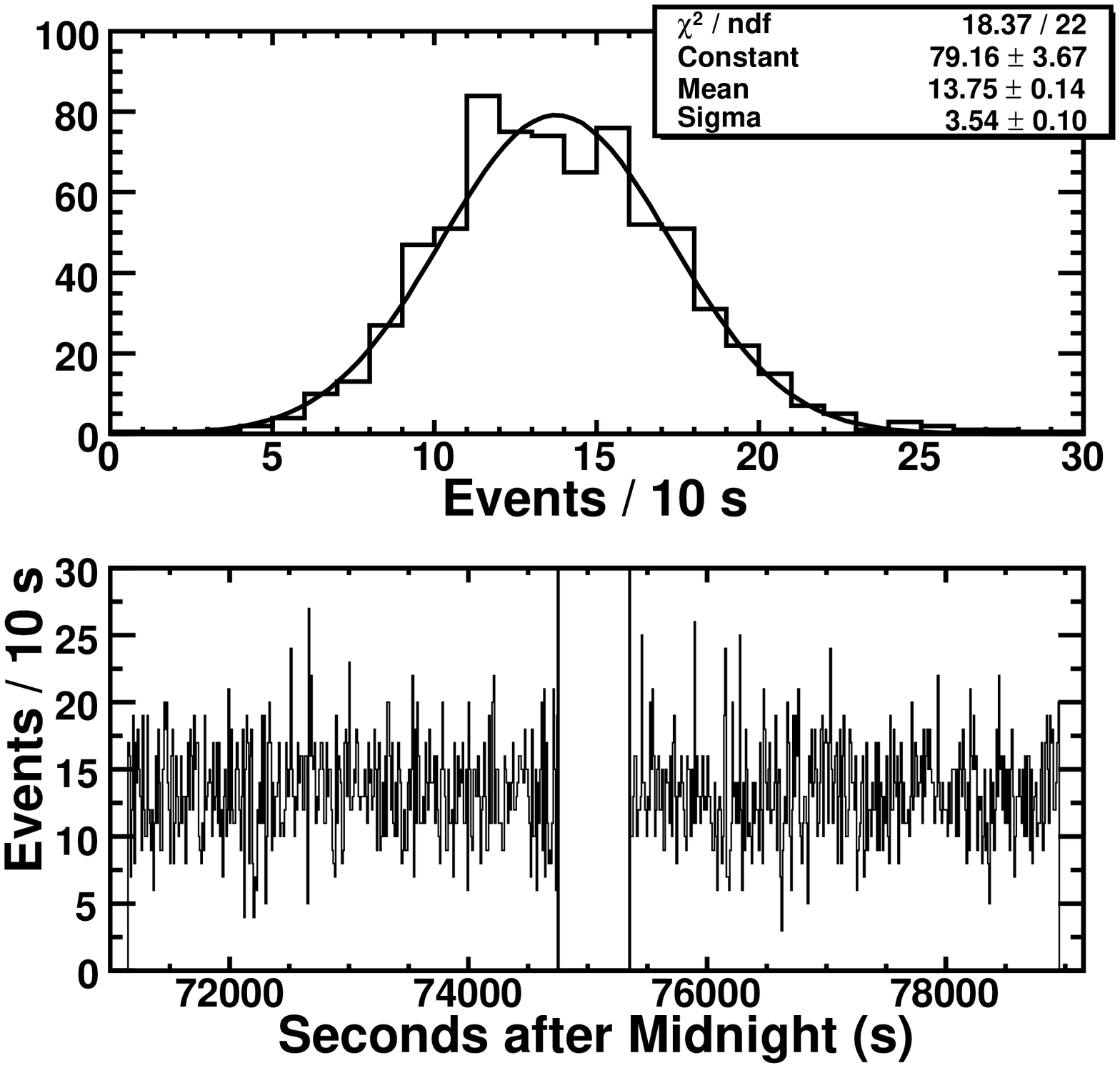}{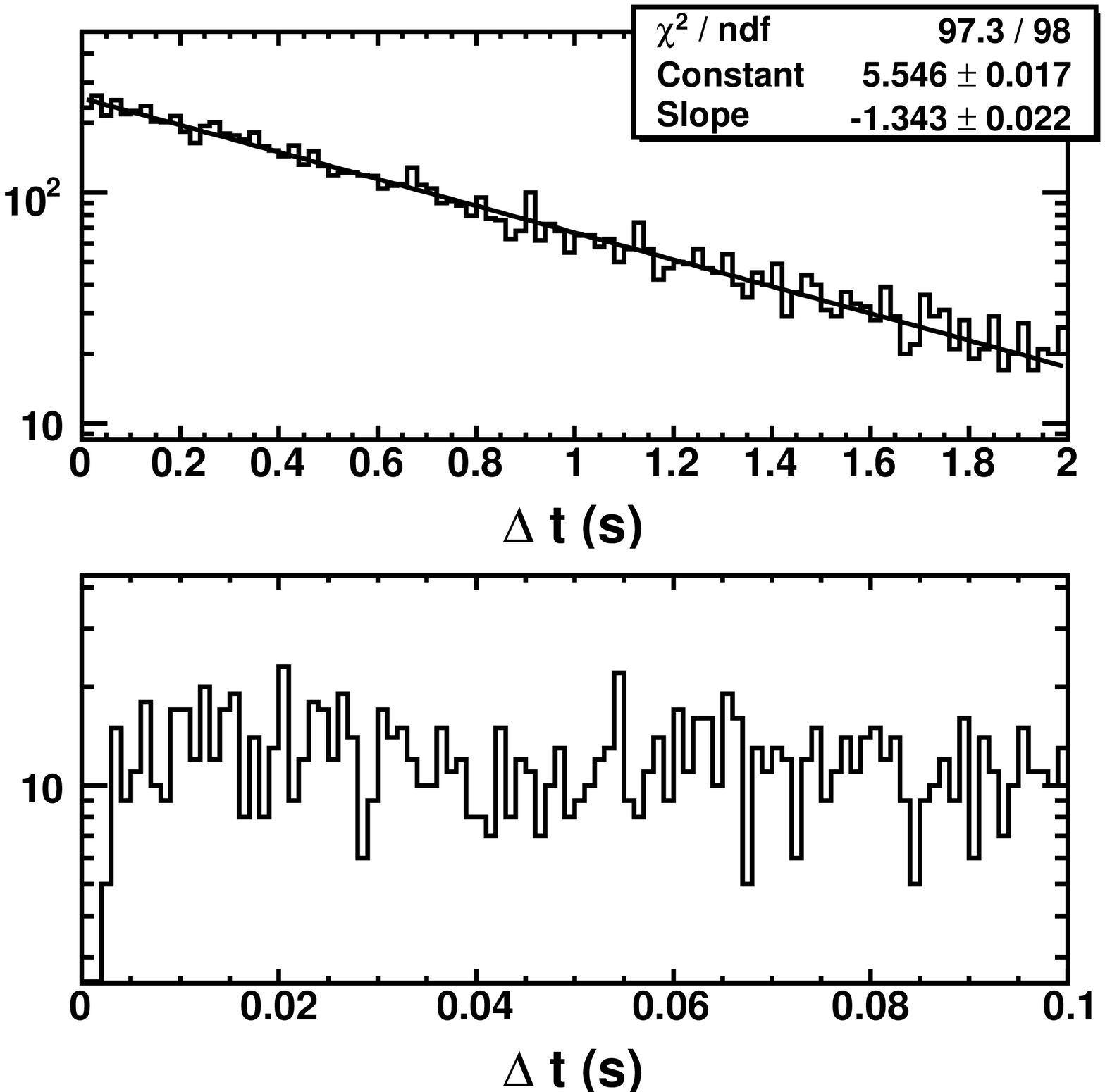}
\caption{\label{fig:trigger_stability} The upper left panel shows the 
  distribution of frequency of events/10s after the filter has been applied
  for GRB000312b. The lower left panel shows events/10s versus time. The gap
  in the middle of the lower left panel corresponds to the on-time window. The
  upper right panel shows the distribution of time difference, $\Delta t$, between 
  consecutive events in the range 0-2~s. The lower right panel is the same as
  the upper right but in the range 0-0.1~s. The gap observed near $\Delta t$=0
  is due to DAQ dead time.}
\end{center}
\end{figure}

\begin{figure}
\begin{center}
\plottwo{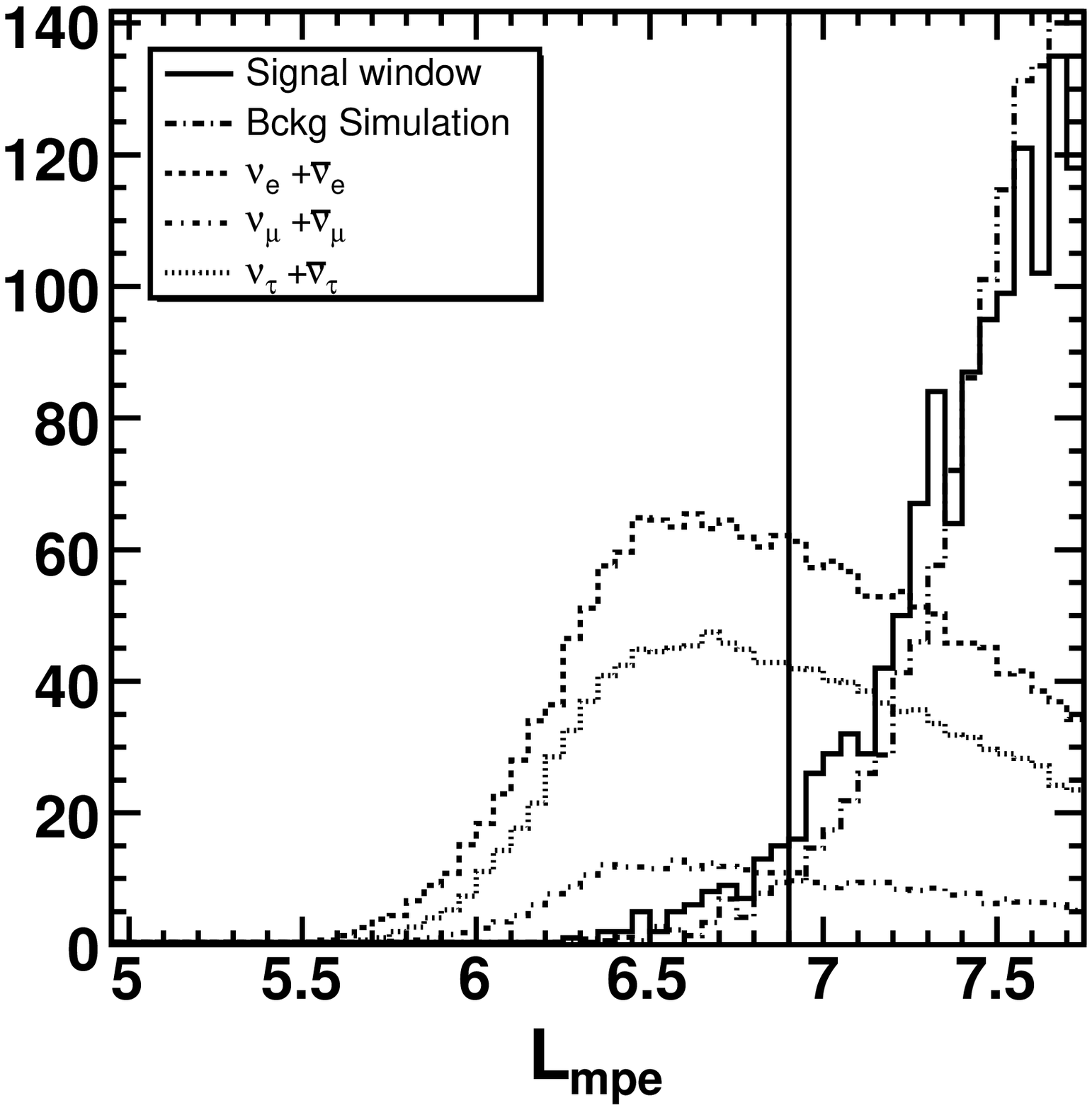}{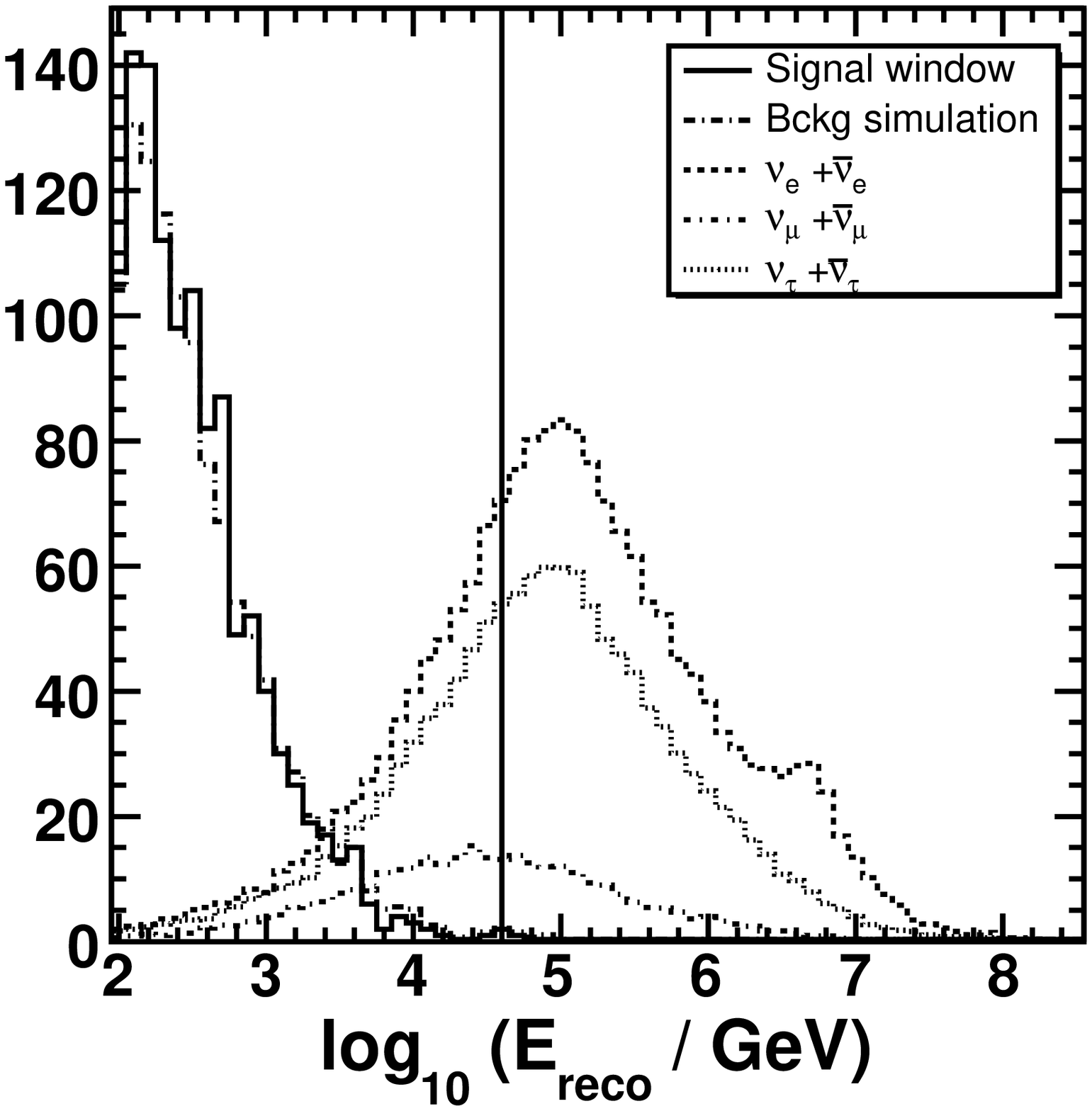}
\caption{\label{fig:filter} The left panel shows the distribution of the
  likelihood parameter, $L_{\mathrm{mpe}}$. Data to the right of the vertical
  line are excluded. The right panel shows the reconstructed cascade energy
  distribution, $E_{\mathrm{c}}$. Data to the left of the vertical line are
  excluded. The signal simulation, following a Waxman-Bahcall spectrum, has
  been scaled up by a factor of 100,000. In both panels the vertical line
  corresponds to the final selection criteria. The background simulation has
  been scaled to match the number of events in the signal window.}
\end{center}
\end{figure}

\begin{figure}
\begin{center}
\plottwo{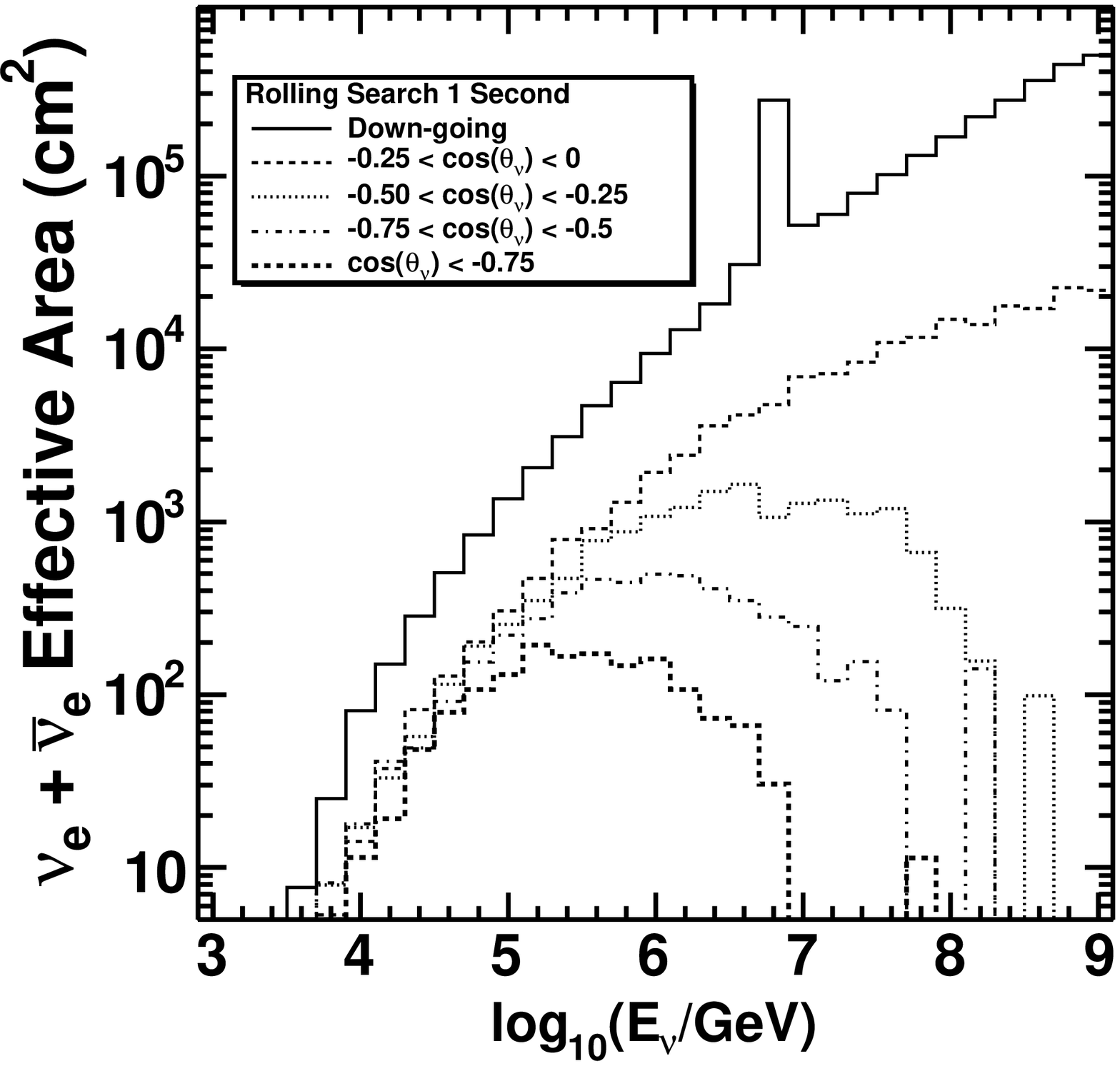}{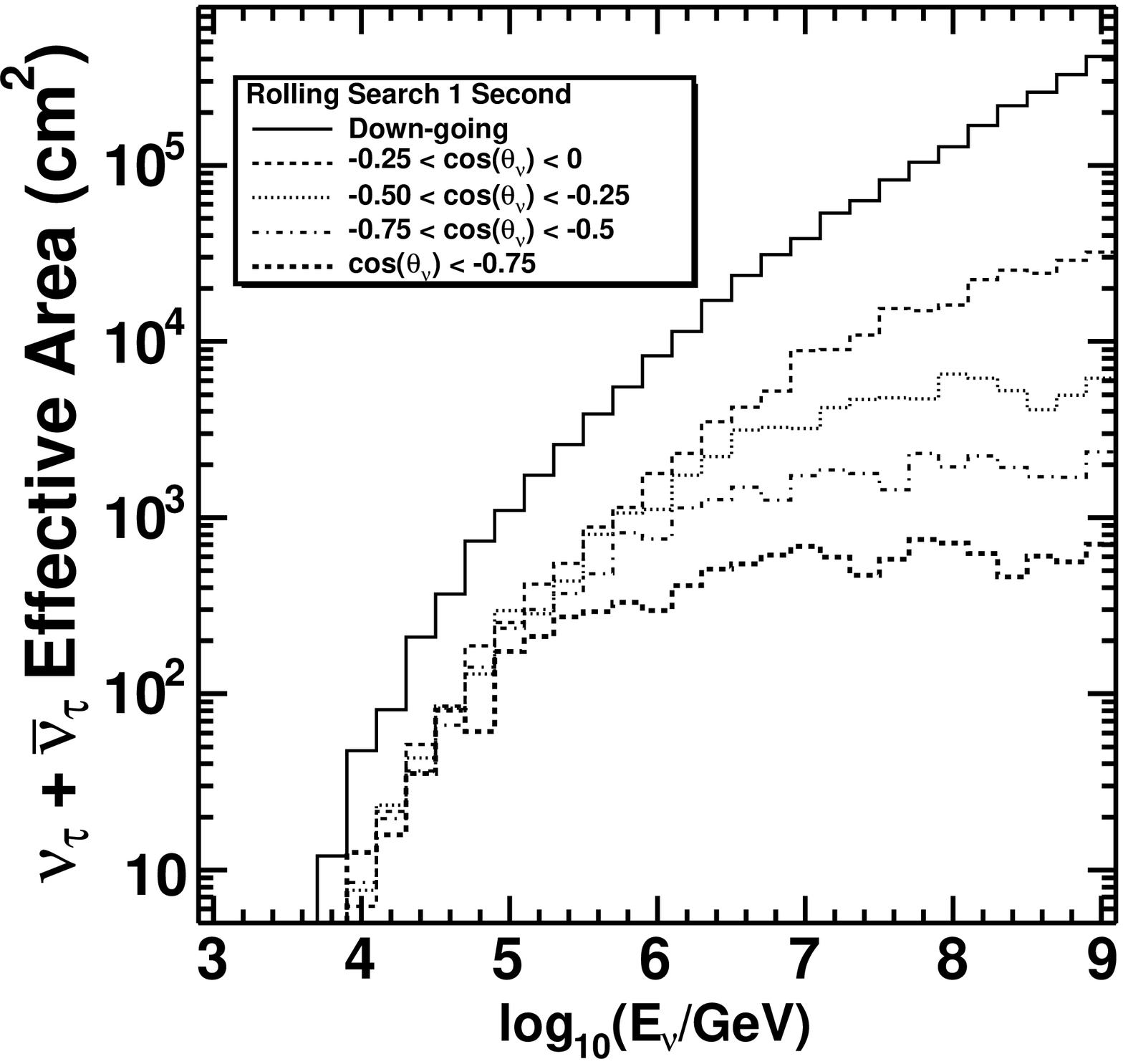}
\plottwo{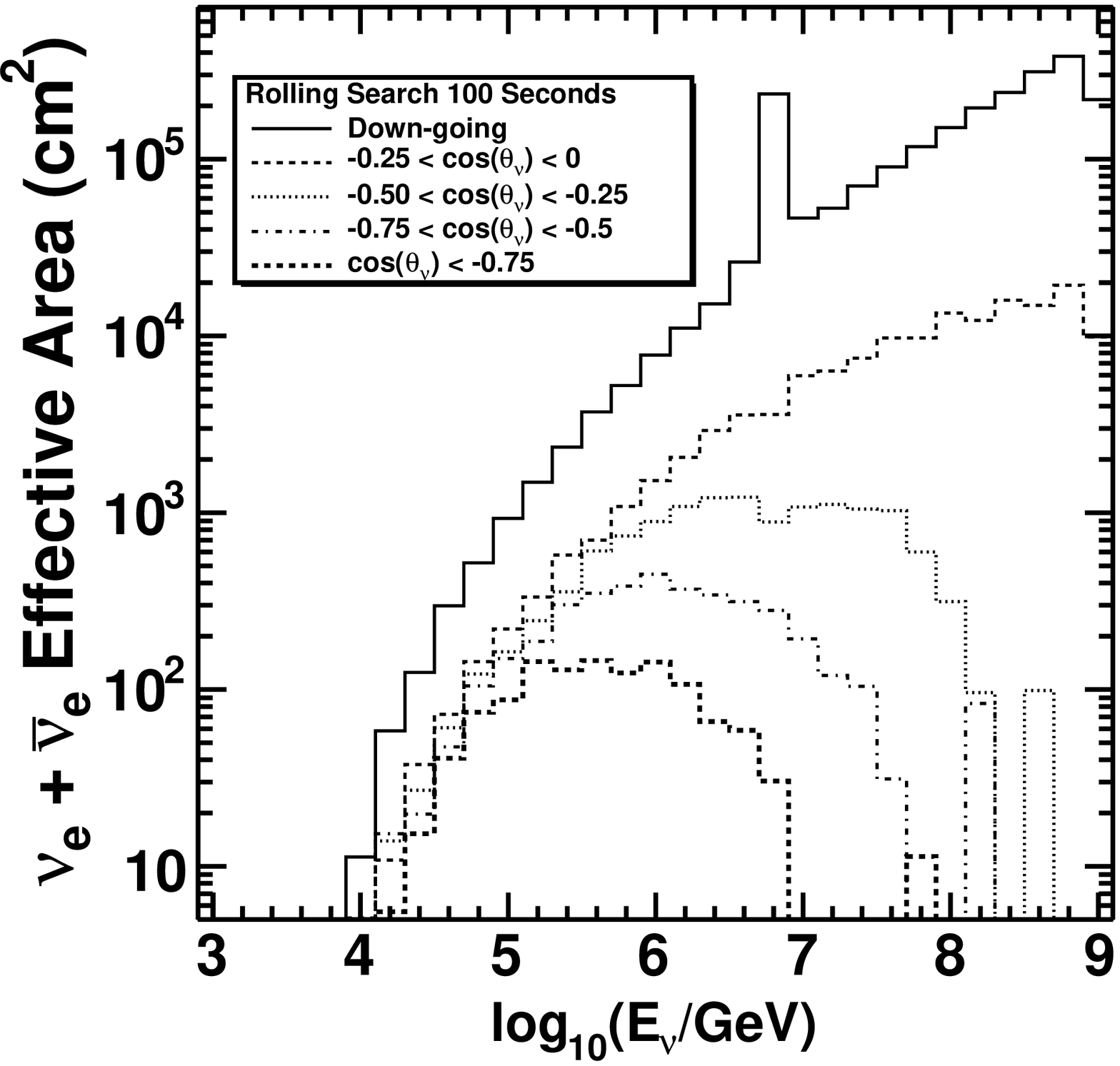}{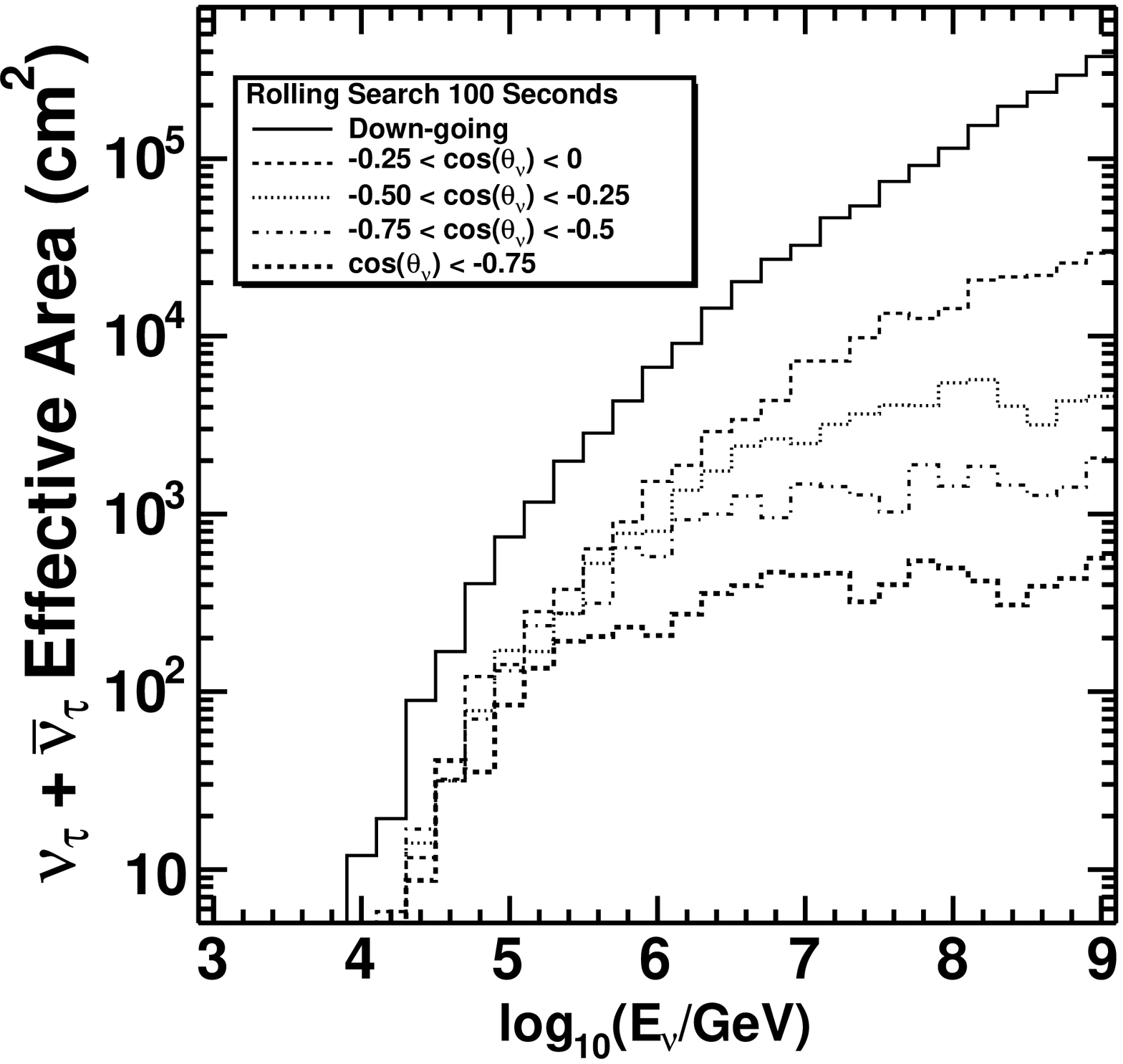}

\caption{\label{fig:effarearoll} Neutrino effective areas as function of 
  neutrino energy (at Earth surface) and $\cos\theta_\nu$ for the rolling
  analysis after all selection criteria have been applied, for both 1 and 100
  second search windows. The peak at 6.3 PeV is due to the Glashow resonance
  for $\bar{\nu}_e$. The effective areas for $\nu_\tau$ for upgoing events are
  larger than for $\nu_e$ because of charged current regeneration. Effective
  areas for $\nu_{\mu}$ and $\bar{\nu}_{\mu}$ are much smaller, because
  neutrino-induced cascades are produced via neutral current interactions only.}
\end{center}
\end{figure}

\begin{figure}
\begin{center}
\plottwo{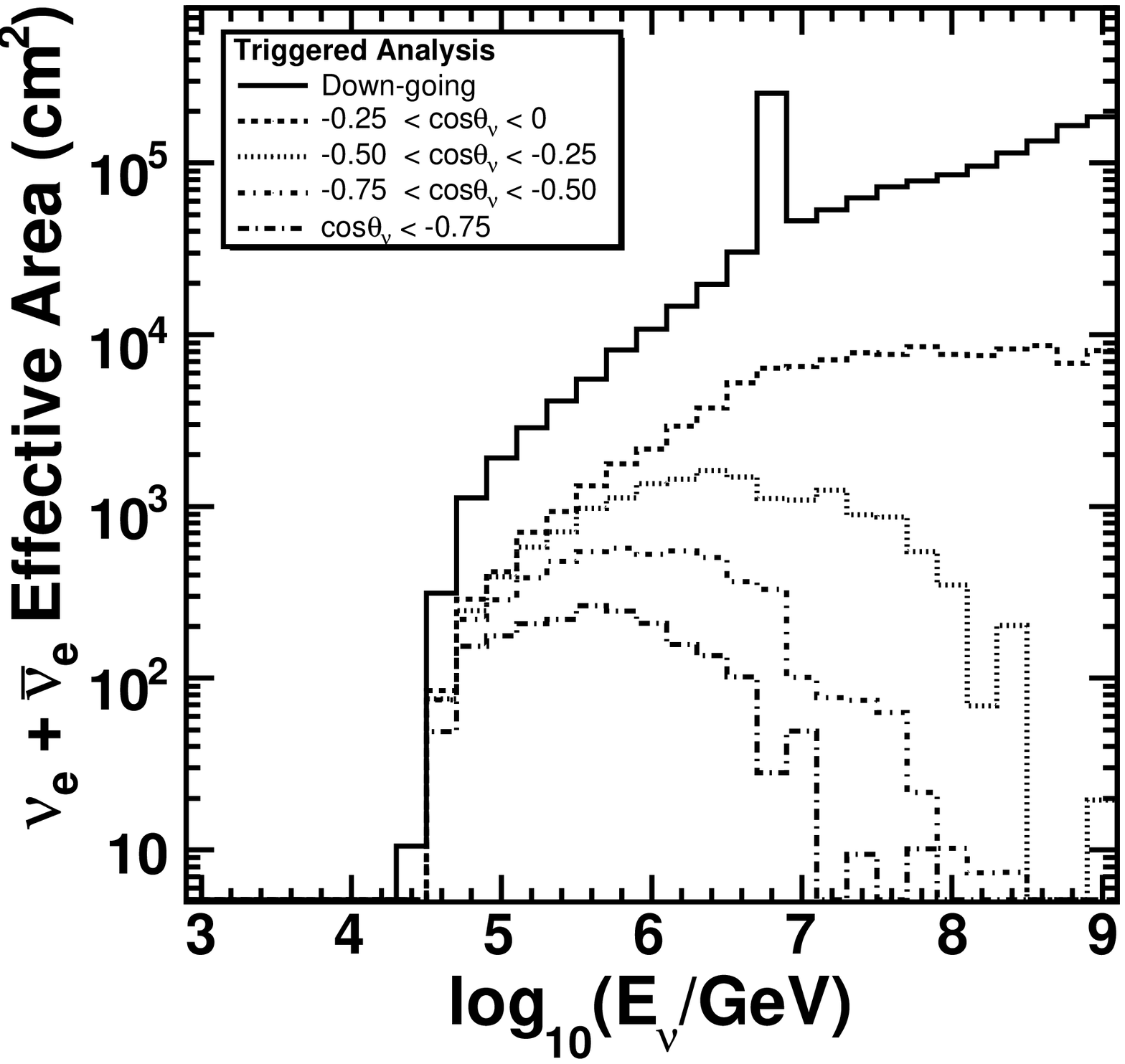}{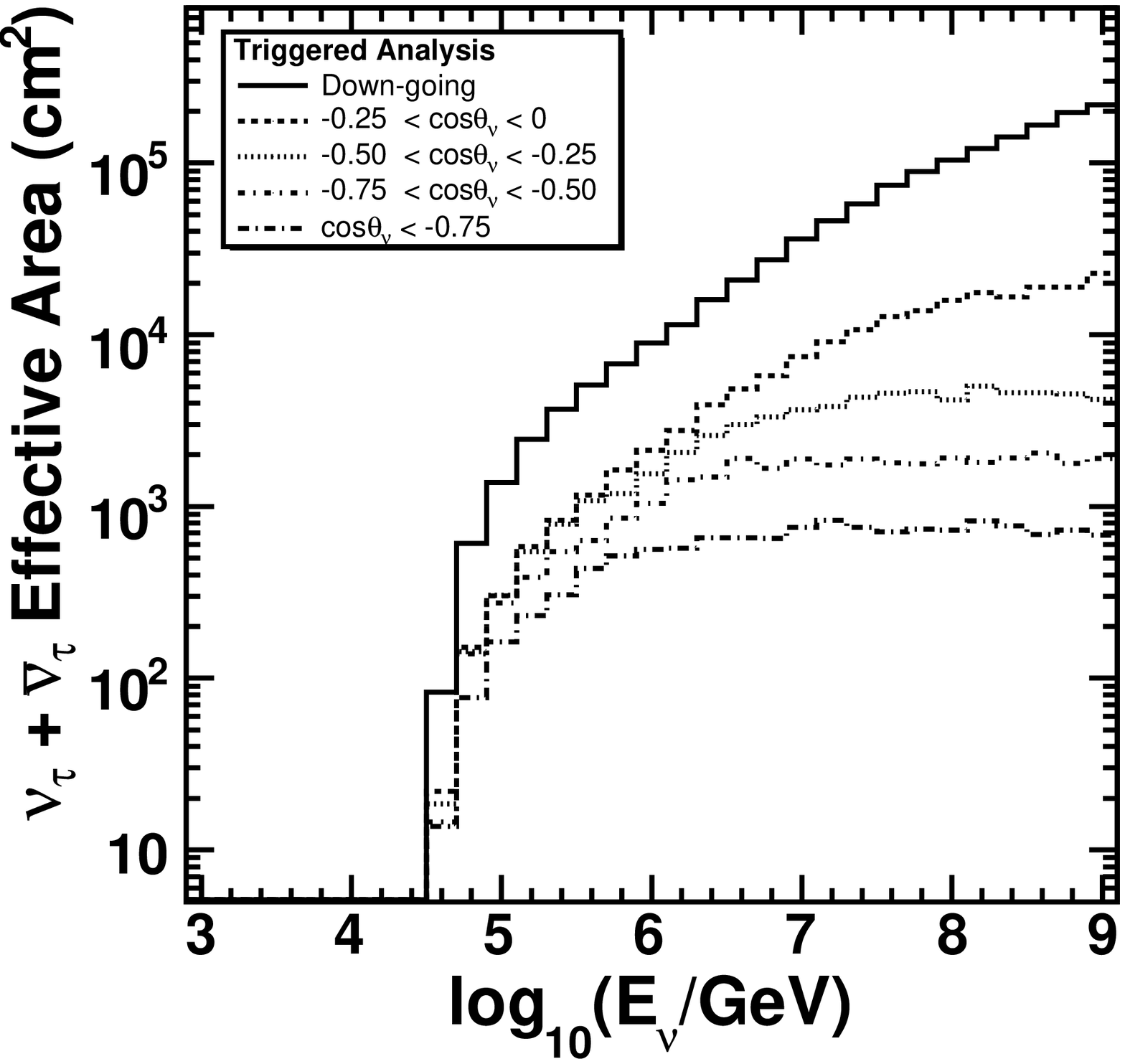}
\caption{\label{fig:effareatrig} Neutrino effective areas as function of
  neutrino energy (at Earth surface) and $\cos\theta_\nu$ for the triggered
  analysis after all selection criteria have been applied. The peak at 6.3 PeV
  is due to the Glashow resonance for $\bar{\nu}_e$. The effective areas for
  $\nu_\tau$ for up-going events are larger than for $\nu_e$ because of charged
  current regeneration.}    
\end{center}
\end{figure}

\begin{figure}
\begin{center}
\plotone{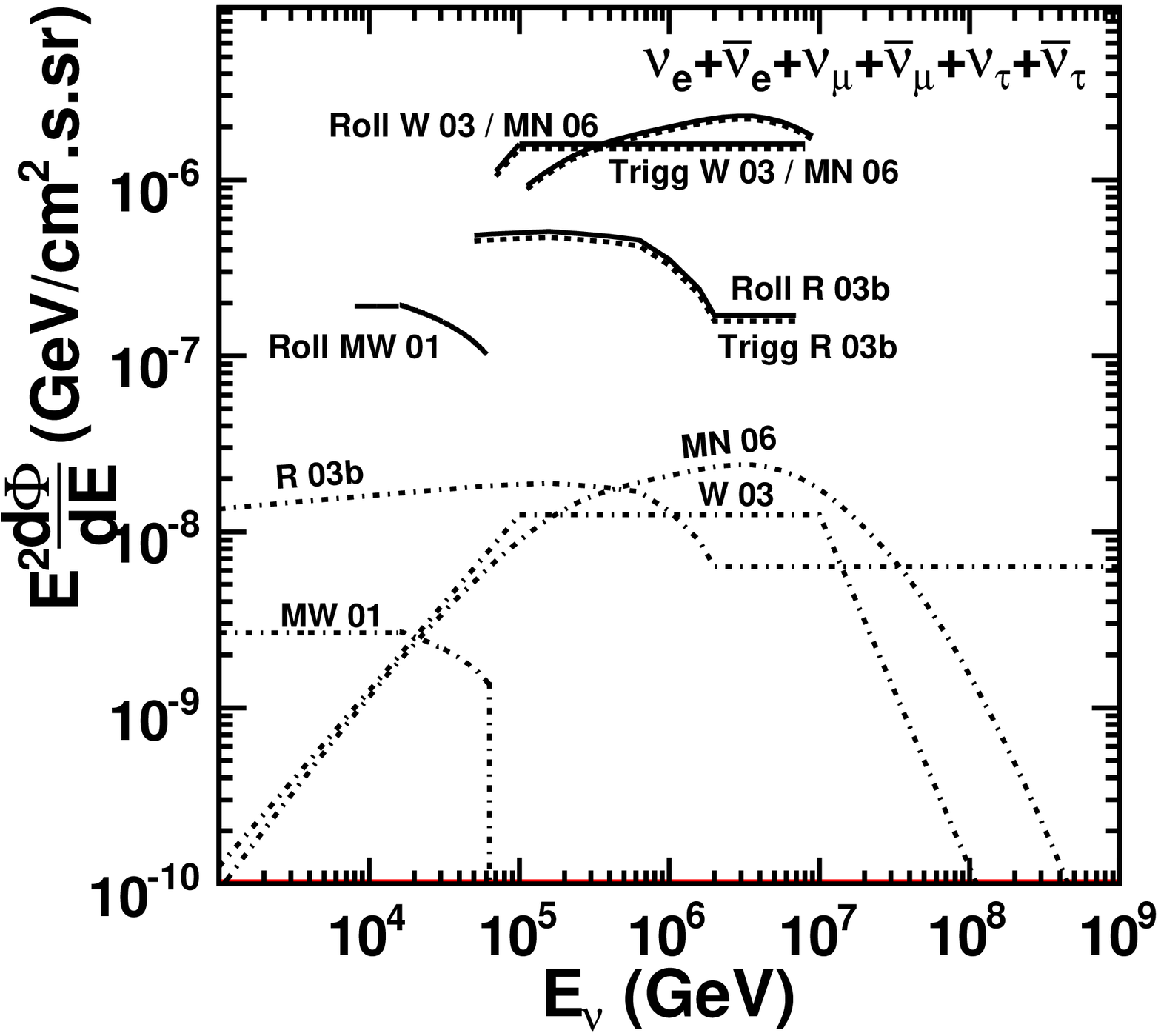}

\caption{\label{fig:limits} Predicted all-flavor diffuse neutrino fluxes and
  experimental limits. Models are shown in dashed-dotted lines: \citet{wb02}
  (Entry W~03, red line in electronic version), \citet{supranova} (Entry
  R~03b, green line in electronic version), \citet{murase} Model A (Entry
  MN~06, blue line in electronic version) and and \citet{choke} (Entry MW~01,
  magenta line in electronic version). All theoretical predictions have been
  adjusted for vacuum oscillations. Also shown are the \textit{Rolling} search
  limits (labeled Roll) in solid line (black in the electronic version), and
  \textit{Triggered} search limits (labeled Trigg) in dashed line (black in the electronic
  version).}
\end{center}
\end{figure}

\end{document}